\documentclass[longauth]{aa}
\usepackage{natbib}
\usepackage{siunitx}
\usepackage{hyperref}
\usepackage{amsmath}
\usepackage{xcolor}
\usepackage{natbib}
\hypersetup{
    colorlinks,
    linkcolor={red!50!black},
    citecolor={blue!50!black},
    urlcolor={blue!80!black}
}

\usepackage{graphicx}
\usepackage{txfonts}
\usepackage{siunitx}

\begin{document} 
   \title{The MAGPI survey: Stellar population radial trends and mass assembly in star-forming galaxies at z$\sim$0.3
   }

   \authorrunning{Mauro et al.}
   \titlerunning{Radial profiles in intermediate-redshift galaxies}

   \author{
      Federica Mauro \inst{1}
      \and
      Bodo Ziegler\inst{1}
      \and
      Iris Breda\inst{1,2}
      \and
      Polychronis Papaderos \inst{2}
      \and
      Gauri Sharma \inst{3,4,5,6,7}
      \and 
      Sabine Thater \inst{1}
      \and 
      Katherine E. Harborne\inst{8,9,10}
      \and
      Trevor Mendel \inst{11,12}
      \and
      Emily Wisnioski\inst{11,12}
      \and
      Claudia Lagos \inst{12,13}
      \and
      Caroline Foster\inst{14}
      \and
      Andrew J. Battisti \inst{13,11,12}
      \and
      Ryan S. Bagge \inst{12,14}
      \and
      Joss Bland-Hawthorn \inst{15}
      \and 
      Scott Croom \inst{15}
      \and
      Siddhartha Gurung-López\inst{16,17}
      \and
      Marcie Mun \inst{18}
      \and
      Tamal Mukherjee \inst{19,20}
      \and
      Sarah M. Sweet \inst{12,21}
      \and 
      Lucas M. Valenzuela \inst{22}
      \and
      Glenn van de Ven \inst{1}
      \and 
      Tayyaba Zafar\inst{19}
   }
   \institute{
     University of Vienna, Department of Astrophysics, Tuerkenschanzstrasse 17, 1180 Vienna, Austria \\ \email{federica.mauro@univie.ac.at} 
     \and
     Instituto de Astrofísica e Ciências do Espaço, Universidade do Porto, Rua das Estrelas, 4150-762 Porto, Portugal
     \and
     University of Strasbourg, CNRS UMR 7550, Observatoire astronomique de Strasbourg, F-67000 Strasbourg, France
     \and
     SISSA- International School for Advanced Studies, Via Bonomea 265, I-34136 Trieste, Italy
     \and 
     UWC- University of the Western Cape, Department of Physics and Astronomy, Cape Town 7535, South Africa
     \and
     IFPU- Institute for Fundamental Physics of the Universe, Via Beirut, 2, 34151 Trieste, Italy
     \and 
     INFN-Sezione di Trieste, via Valerio 2, I-34127 Trieste, Italy
     \and
     Institute for Computational Cosmology, Durham University, South Road, Durham DH1 3LE, UK 
     \and 
     Centre for Extragalactic Astronomy, Durham University, South Road, Durham DH1 3LE, UK 
     \and
     Department of Physics, Durham University, South Road, Durham DH1 3LE, UK
     \and
     Research School of Astronomy and Astrophysics, Australian National University, Cotter Road, Weston Creek, ACT, 2611, Australia
     \and 
     ARC Centre of Excellence for All Sky Astrophysics in 3 Dimensions (ASTRO 3D)
     \and
    International Centre for Radio Astronomy Research, The University of Western Australia, 35 Stirling Highway, Crawley WA 6009, Australia
    \and
     School of Physics, University of New South Wales, Sydney, NSW 2052, Australia
     \and 
     Sydney Institute for Astronomy, School of Physics, University of Sydney, NSW 2006, Australia
     \and 
     Observatori Astron\`omic de la Universitat de Val\`encia, Ed. Instituts d’Investigaci\'o, Parc Cient\'ific. C/ Catedr\'atico Jos\'e Beltr\'an, n2, 46980 Paterna, Valencia, Spain
     \and 
     Departament d’Astronomia i Astrof\'isica, Universitat de Val\`encia, 46100 Burjassot, Spain
     \and 
     Institut d'Astrophysique de Paris, UMR 7095, CNRS, Sorbonne Université, 98 bis boulevard Arago, 75014 Paris, France
     \and
     School of Mathematical and Physical Sciences, Macquarie University, NSW 2109, Australia
     \and 
     Astrophysics and Space Technologies Research Centre,  Macquarie University, Sydney, NSW 2109, Australia
     \and 
     School of Mathematics and Physics, University of Queensland, Brisbane, QLD 4072, Australia
     \and
     Universitäts-Sternwarte, Fakultät für Physik, Ludwig-Maximilians-Universität München, Scheinerstr. 1, 81679 München, Germany
     }

   \date{Received XXX; accepted XXX}

  \abstract{The evolution of galaxies from cosmic noon to the present day provides a key window through which to probe the balance between early, rapid bulge formation and prolonged disk growth. In this context, the epoch at $z \sim 0.3$ (intermediate redshift) marks a crucial transitional phase between the peak of cosmic star formation and the predominantly quiescent local Universe, where these processes can be directly observed.
  }{In this work, we examine the spatially resolved stellar populations of 34 galaxies at $z \sim 0.3$ to quantify radial gradients in age, stellar metallicity, and star formation activity, and disentangle the distinct evolutionary pathways of inner and outer galactic components.}{We utilised MUSE integral-field spectroscopy data cubes from the MAGPI survey at redshifts of $0.28 < z < 0.35$. Stellar population properties were derived using the spectral synthesis codes \texttt{FADO} and \texttt{Starlight}, and radial profiles were constructed by fitting isophotal annuli to the galaxy continuum emission, thereby preserving galaxy morphology and enabling the separation between components. We further reconstructed star formation histories and cumulative mass assembly curves for inner and outer regions.}{We find pronounced negative radial gradients in age and negative to flat gradients in stellar metallicity. Inner regions are systematically older and more metal-rich than their surrounding outskirts, with age differences of up to 3-4 Gyr in the most massive systems. H$\alpha$ equivalent width profiles reveal centrally suppressed specific star formation in most galaxies, consistent with inside-out quenching. Star formation histories and mass assembly curves demonstrate that galaxy cores formed $80\%$ of their stellar mass rapidly, within the first 2-3 Gyr of cosmic time; while areas outside  $\mathrm{1\,R_{eff}}$ assembled more gradually and sustained star formation to later epochs. These findings support a scenario in which early dissipative processes and mergers dominate inner formation. At the same time, outskirts evolve primarily through extended, secular star formation, establishing the centrally concentrated quenching and inside-out growth that link high-redshift systems to the quiescent galaxies of the local Universe.}{}


   \maketitle

\section{Introduction} \label{sec:intro}
The hierarchical assembly of galaxies within the $\Lambda$ cold dark matter ($\Lambda$CDM) cosmological framework provides a robust basis for understanding the growth of structure in the Universe. Yet, the baryonic physics shaping the galaxies we observe today remains a major frontier of extragalactic astronomy \citep{Somerville_2015}. Over the past three decades, advances in observational capabilities, in particular the advent of large-class telescopes, space-based imaging from the Hubble Space Telescope (HST) and the James Webb Space Telescope (JWST), and integral field spectrographs, have transformed our ability to investigate galaxy evolution across cosmic time. As a result, fundamental scaling relations such as the Tully-Fisher relation \citep{Tully, ziegler2002, Bohn_2004, Sharma_2024} and the stellar mass-metallicity relation \citep{Mannucci_2010,Henry_2013, Maier2015, Maier2019, looser2024} have been traced over a wide redshift range, extending to early cosmic epochs, revealing a downsizing, or anti-hierarchical, trend in galaxy star formation histories \citep[SFHs;][]{Thomas_2010,Tonini_2011, Bohn2016, Perez_2021}.

A typical feature of Milky Way-like galaxies is the co-existence of two main structural components: a centrally concentrated bulge and a more extended, rotationally supported exponential disk \citep{Freeman1970}. In the local Universe, large integral field spectroscopy (IFS) surveys such as CALIFA \citep{Sanchez2012}, MaNGA \citep{Bundy2015}, and SAMI \citep{Croom2012} have enabled detailed archaeological approaches to study these components. These efforts combine spectro-photometric bulge-disk decomposition \citep{MendezAbreu2021, Costantin2022} with dynamical modelling techniques \citep{Jethwa2020, Thater2023} that explicitly account for the orbital structure and mass distribution of galaxies. Early work by \citet{Zhu_2017} demonstrated that bulge-disk decomposition can be performed robustly within a fully dynamical framework. Subsequent studies have extended this approach by coupling dynamical models with stellar population synthesis, allowing for the simultaneous recovery of kinematic, structural, and stellar population properties of galaxy components (e.g. \citealt{Poci2021, Thater_2023}). 

These studies show that bulge properties, in particular stellar ages and metallicities, form a continuum that correlates with those of their host disks \citep{Breda_2018}. This supports a unified bulge-disk coevolution framework, in which the relative importance of early, rapid central formation and slower, secular disk growth varies systematically with galaxy mass. Consistently, gradients in stellar age and metallicity studies \citep{Perez2013,Lu2023}, together with various works on observational features \citep{Nelson_2016,Spilker_2019,Smith_2021}, indicate that disk galaxies have predominantly grown inside-out.

Extending such spatially resolved studies beyond the nearby Universe remains challenging. A major limitation of most high-redshift investigations is that they rely primarily on global, and thus luminosity-weighted properties that average over potentially large internal differences between inner and outer stellar populations. Because these regions follow distinct star-formation (SF) and chemical-enrichment histories, they also exhibit different spectral energy distributions (SEDs) and mass-to-light ratios \citep{Tortora_2011,Garcia_2019,Breda_2023}. Central quenching of SF can further bias photometric bulge-disk decompositions by oversubtracting disk light within the bulge radius, leading to systematic underestimates of bulge luminosities and stellar masses \citep{Pap22,POB23}. More generally, SED variation, for example, a red, quenched bulge embedded in a blue, star-forming disk, complicates the interpretation of unresolved measurements and their comparison across redshift.

While the local Universe provides a high-fidelity view of the end products of galaxy evolution, it is at intermediate and high redshift that the mechanisms of mass assembly and quenching can be directly probed. Near the peak of cosmic SF at $z \sim 1-2$, galaxies often exhibit rapid and spatially extended growth, and quenching appears to act on global scales \citep{Costantin2019, Lu2023, Mun2024}. In contrast, present-day massive disk galaxies predominantly show centrally suppressed SF. A key open question is therefore when and how the dominant mode of quenching transitioned from being global to being centrally concentrated.

The redshift regime around $z \sim 0.3$, corresponding to a look-back time of roughly $3-4\,$ Gyr, represents a crucial yet relatively unexplored phase in this transition. It occurs after the decline from the cosmic SF peak, but before galaxies settled into their present-day, largely quiescent state \citep{Fritz2014}. This epoch is ideally suited to capturing the emergence of centrally driven processes such as nuclear suppression of SF and inside-out quenching, while significant SF is still ongoing in galaxy disks.

In this work, we use data from the \textit{Middle Ages Galaxy Properties with Integral Field Spectroscopy} (MAGPI) survey \citep{Foster2021} to investigate the spatially resolved stellar populations of star-forming galaxies at $z \sim 0.3$. In the absence of high-resolution space-based photometric imaging for the full sample, we adopt the circularised effective radius, $R_{\rm eff}$, as a physically motivated scale to separate inner and outer regions. While this approach necessarily simplifies the diversity of galaxy structures, it provides a consistent framework for comparing stellar population trends across the sample and how it relates to other similar studies, both at intermediate and local redshift regimes.

Stellar population properties were derived through spectral synthesis of the MUSE data cubes using the \texttt{FADO} and \texttt{Starlight} codes. This analysis yielded spatially resolved maps of stellar ages, metallicities, extinction, mass surface density, and SFHs. To robustly characterise radial trends while preserving galaxy morphology, we constructed profiles using isophotal annuli (\textsc{isan}). In addition to age and metallicity gradients, we analysed H$\alpha$ equivalent-width profiles and traced the spatial distribution of ongoing SF, and we reconstructed cumulative mass assembly curves separately for inner and outer regions.

The aim of this paper is to quantify the radial behaviour of stellar age and metallicity in massive star-forming galaxies at $z \sim 0.3$, and to compare the evolutionary pathways of their inner and outer regions. By linking spatially resolved stellar population gradients, SF indicators, and mass assembly histories, we assess whether galaxies at this epoch already exhibit clear signatures of centrally concentrated growth and inside-out quenching. In doing so, this work bridges the gap between the globally active systems observed near cosmic noon and the predominantly inside-out evolutionary patterns characteristic of disk galaxies in the local Universe.

In this paper, Section \ref{data} describes the survey and sample selection. Section \ref{method} outlines the methods used to derive the physical quantities with spectral synthesis tools. The main results are presented in Section \ref{results}, with a particular focus on radial profiles and the properties of inner and outer regions. Finally, the findings and mass assembly histories are discussed and summarised in Sections \ref{discussion}-\ref{summary}. Throughout the work we adopt a flat $\Lambda$CDM cosmology with $H_0 = 70\,\mathrm{km\,s^{-1}\,Mpc^{-1}}$, $\Omega_m = 0.3$, and $\Omega_\Lambda = 0.7$.

\section{Data sample}
\label{data}
\subsection{The MAGPI survey}
\label{sec::magpi}
The MAGPI survey \citep{Foster2021} is a Large Program on the VLT using MUSE to map the spatially resolved stellar and ionised gas properties of galaxies at intermediate redshifts, corresponding to look-back times of $\sim 3-4$ Gyr. This epoch represents a key transitional phase between the dynamically active high-redshift Universe and the more quiescent present day, providing a crucial link in the study of the co-evolution of baryonic and dark matter \citep{Santucci_2024,Sharma2026}, morphological transformation and properties \citep{Deugenio_2023a, Deugenio_2023b}, and environmental processing \citep{Derkenne2023,Derkenne_2024a, Derkenne_2024b, Foster_2025, Mun_2025}.
The survey targets 60 massive primary galaxies with $M_\star > 7\times 10^{10} \ \mathrm{M}_\odot$, selected mainly from the Galaxy and Mass Assembly (GAMA) survey \citep{Driver2011}, and spanning a wide range of environments from isolated systems to rich groups (halo masses $\log (M_{\mathrm{halo}}/\mathrm{M}_\odot) \approx 12-15$) (\cite{Foster_2025}, Harborne et al. in prep).
In addition, each $1'\times 1'$ MUSE field captures a substantial number of secondary galaxies, greatly enriching the mass and environmental coverage. Observations, performed with MUSE in Wide Field Mode, deliver a spectral range of 4700-9350\,\AA\ at 1.25\,\AA\ sampling \citep{Foster2021}.

\subsection{Galaxy selection}
\begin{figure} 
    \centering
    \includegraphics[scale = 0.45]{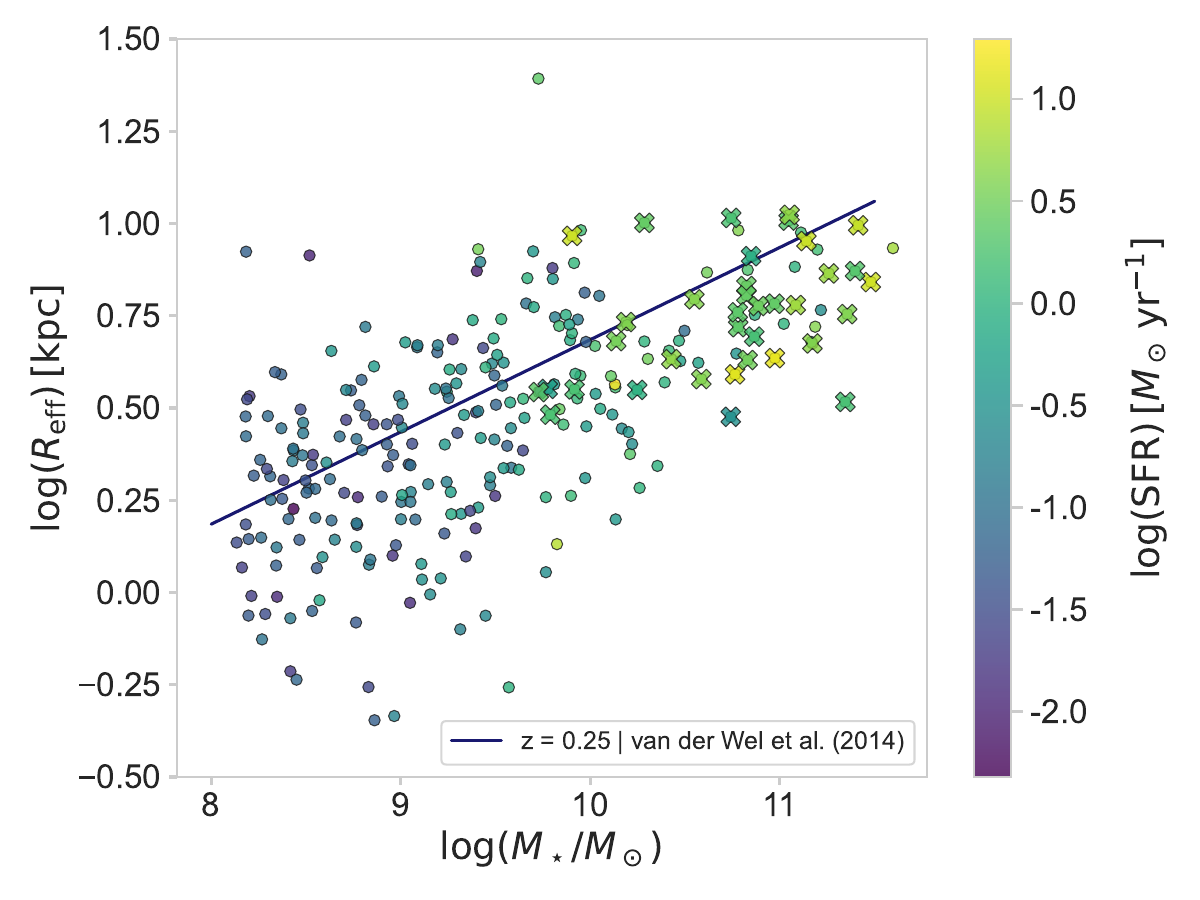}
    \caption{Mass-size relation of H$\alpha$ strong emitters MAGPI galaxies, colour-coded by SFR. $\log(M_{\star}/M_{\odot})$, $\log R_{\rm eff}$, and $\log(\rm SFR)$ are taken from MAGPI \textsc{ProSpect}, \textsc{ProFit}, and line flux measurement catalogues, respectively. The sample (dots) has been taken from \cite{Sharma2026}, and the crosses represent the galaxies included in this work. The solid line represents the mass-size relation for galaxies at $z=0.25$ from \cite{van_der_Wel_2014}. We note that the sample, while centred around $z\sim 0.3$, extends over the redshift interval $0.1\leq z\leq 0.42$}
    \label{fig:mass_size}
\end{figure}

From the MAGPI parent catalogue, we selected a subsample of 34 galaxies, optimised for spatially resolved spectral analysis and structural decomposition (Fig.~\ref{fig:mass_size}). Central to our selection was the requirement of robust H$\alpha$ emission up to at least $2 \, R_{\rm eff}$, ensuring sensitivity to ongoing SF across different galaxy regions. We selected galaxies confirmed to exhibit significant H$\alpha$ flux ($S/N > 4$), ensuring that the sample includes galaxies with stellar populations that are not fully quenched yet still display strong nebular signatures. Most galaxies in our sample are also included in works by \cite{Koller2024} and \cite{Sharma2026}, based on strong optical emission-line selection criteria for SFR, gas-phase metallicities, and spatially resolved kinematic analyses.
We acknowledge that this procedure naturally biases the sample towards more disk-like and more extended systems. However, such morphologies are essential for our work, as they permit a meaningful investigation of stellar-population properties across distinct isophotal regions.

The redshift range was restricted to $0.28 < z < 0.35$ to provide uniform coverage of H$\beta$, [O\,\textsc{iii}]$\lambda5007$, H$\alpha$, [N\,\textsc{ii}]$\lambda6583$, [O\,\textsc{ii}]$\lambda3727$, and [S\,\textsc{ii}]$\lambda\lambda6717,6731$. To ensure sufficient spatial resolution, galaxies with effective radii in the $i$ band smaller than the $0.7$ \unit{\arcsecond} have been excluded. Systems with axis ratio $b/a < 0.35$ were removed to minimise projection effects and dust attenuation, following the same criteria as \cite{Koller2024}.
This selection therefore favours relatively regular, extended systems, typically at the high-mass end ($M_{*} \geq 10^{10.5} \, M_{\odot}$), for which distinct structural components can be separated and analysed.

The final sample of 34 galaxies is shown in Fig.~\ref{fig:mass_size} in the stellar mass versus effective radius plane. Stellar masses shown are taken from MAGPI's \textsc{ProSpect} \citep{Robotham_2020} catalogue, while effective radii come from \textsc{ProFit} catalogues \citep{Robotham2017, pro} and star formation rates from Balmer line flux ratio measurements \citep{Battisti_2026}.

\section{Methodology}
\label{method}
\subsection{Data preparation}
\label{sec:data_prep}

Processing and preparation of the MUSE data cubes were carried out using the \texttt{GLANCE} (Galactic archaeoLogy via chronochemicAL \& dyNamiCal modElling\footnote{\url{https://gitlab.com/iris.b/glance}}; \cite{Breda2026}) pipeline, a fully automated Python-based framework designed for photometric, spectral, and dynamical analysis of galaxy data cubes. The pipeline can also be used to perform flux calibration, galactic foreground extinction correction, and rest-frame transformation. Additionally, \texttt{GLANCE} includes routines for masking foreground stars and unrelated background sources.

To preserve the structural and morphological information of each galaxy, we selected spaxels with the highest signal-to-noise ratio (S/N), ensuring wide spatial coverage across morphological components. A Voronoi binning scheme was applied using the algorithm by \cite{cappellari2003}, targeting a S/N of 8 in the rest-frame $6050-7500$  \AA\ region, where the instrument has peak sensitivity. This typically yielded $\sim80-300$ bins per galaxy, depending on the galaxy's extent and surface brightness profile.

Foreground extinction was corrected using the \texttt{irsadust} Python package, which applies the \cite{Cardelli1989} extinction law with $R_V = 3.1$ and extinction values from \cite{Schlafly_2011}. All spectra were then de-redshifted using the spectroscopic redshift estimate.

\subsection{Stellar population synthesis}
Each Voronoi-integrated, rest-frame spectrum was analysed with two spectral synthesis codes: \texttt{FADO} \citep{Gomes_2017} and \texttt{Starlight} \citep{Star}. 
\texttt{FADO} simultaneously models both stellar and nebular continuum emission in a fully self-consistent manner. By enforcing energy balance between the stellar ionising continuum and the nebular line fluxes, \texttt{FADO} can more accurately recover the physical properties of star-forming regions, especially in spectra with significant nebular contamination. In this regime, purely stellar methods tend to bias age and metallicity estimates \citep{Byler_2017,Cardoso_2019}. On the other hand, \texttt{Starlight} lacks such a self-consistency mechanism. Both \texttt{FADO} and \texttt{Starlight} decompose the stellar emission into a linear combination of SSPs convolved with a Gaussian function to model the stellar velocity dispersion. While it does not account for nebular emission, \texttt{Starlight} has been extensively validated for studying evolved stellar populations and serves here as a benchmark for assessing the impact of nebular continuum treatment on derived stellar properties \citep{Cardoso_2022, Miranda_2025}. 
Gaussian models were fitted to the principal emission lines when present, including H$\beta$, [O\,\textsc{iii}]  $\lambda\lambda4959,5007$, [N\,\textsc{ii}]  $\lambda\lambda6548,6584$, H$\alpha$, and [S\,\textsc{ii}]  $\lambda\lambda6717,6731$. Fits were performed for each Voronoi bin, using initial conditions from the continuum-subtracted spectra.

To correct for internal extinction, we used the observed Balmer decrement, H$\alpha$/H$\beta$, assuming an intrinsic value of 2.86 and a \cite{Cardelli1989} extinction curve. These corrections were propagated to all measured emission-line fluxes.

Both \texttt{FADO} and \texttt{Starlight} were run with identical SSP libraries based on the updated \cite{BC03} models and a \cite{Chabrier_2003} initial mass function, covering ages from $1$ Myr to $10$ Gyr and four metallicities ($Z = 0.001$ to $Z = 0.0191$). The SSP library was truncated to the lookback time corresponding to the galaxies redshift ($z \sim 0.3$), thereby excluding templates older than the age of the Universe at the observed epoch. The resulting fits yielded maps of stellar mass, light and mass-weighted stellar age and metallicity, together with emission-line fluxes and equivalent widths. These were compiled into FITS datacubes, where each layer corresponds to a different derived quantity. We also extracted the mass assembly histories for all age and metallicity bins in the stellar library, for each Voronoi bin of our galaxies.
We refer to  Appendix \ref{app:FS} for a direct comparison of properties extracted with \texttt{FADO} and \texttt{Starlight}.

\subsection{Isophotal annuli}
\begin{figure}
\centering
    \includegraphics[scale = 0.4]{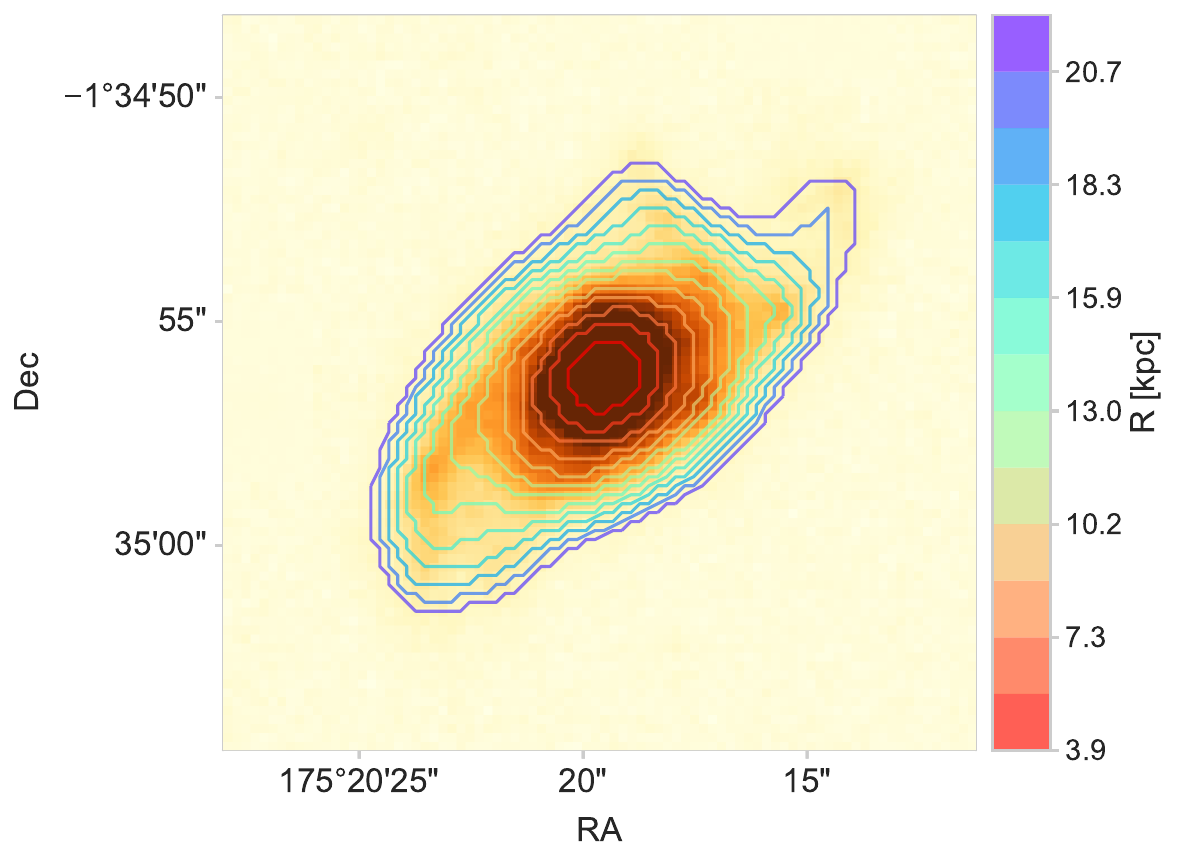}
    \includegraphics[scale =0.35]{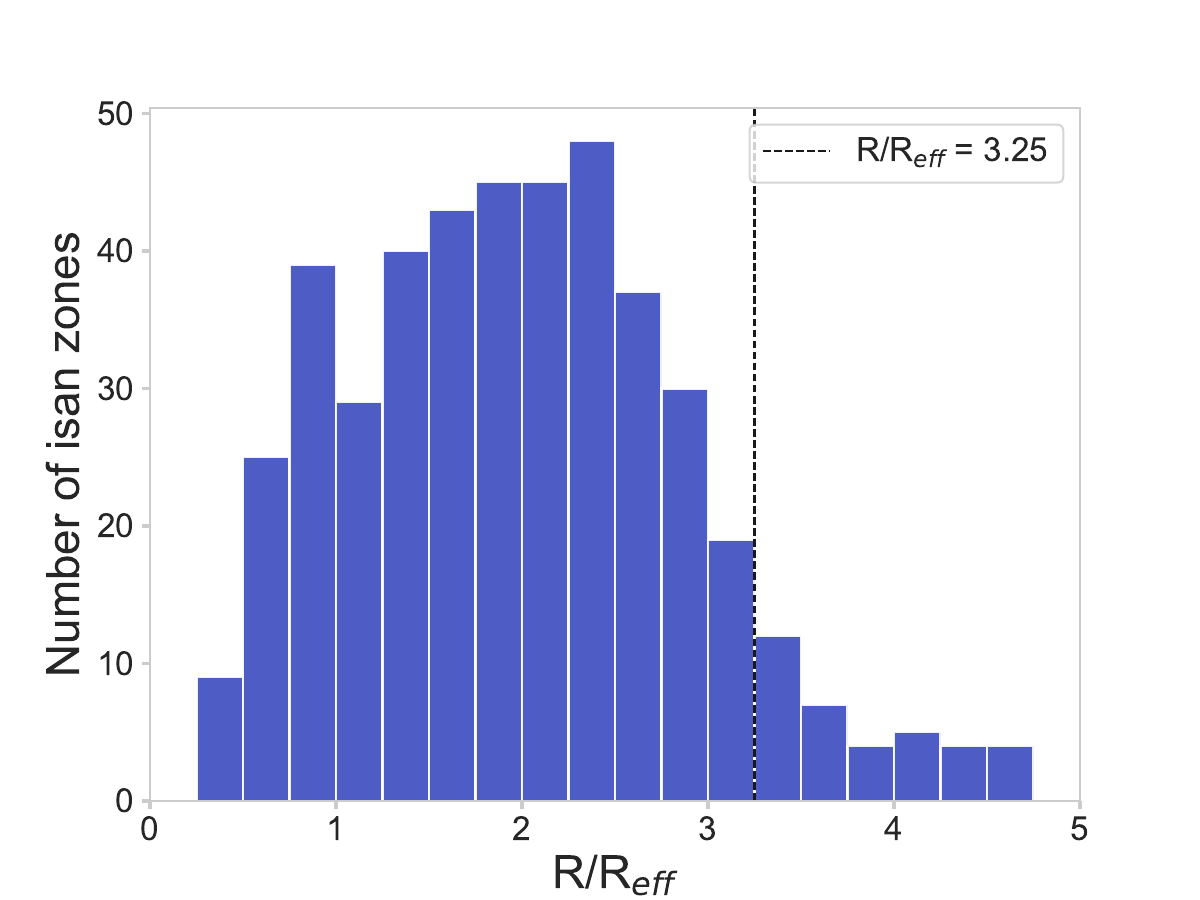}
    \caption{Isophotal zones. The top panel shows the continuum map of galaxy MAGPI 1202197197 with the equidistant isophotal zones (contours), obtained through the \textsc{isan} technique, colour-coded from red to purple for growing circularised $\rm R_{\rm eff}$ of each isophotal zone. The colour bar represents the radius corresponding to each  \textsc{isan} zone (kiloparsecs). The galaxy morphology is traced by each contour, which includes the spaxels associated with the corresponding zone. The bottom panel histogram illustrates the total number of isophotal zones for the entire galaxy sample, as a function of $\rm R/R_{\mathrm{eff}}$. The vertical black line is placed at $\rm R/R_{\mathrm{eff}}$ = 3.25, the upper radius threshold for this study for galaxies in the high mass regime of $11 \leq \log(M_{\star}/M_{\odot}) \leq 11.5$.}
    \label{fig:isan}
\end{figure}

To derive spatially resolved radial trends of the stellar population properties, we employed the \textsc{isan} surface photometry technique originally developed by \cite{Papaderos2002} and adapted in similar spectral synthesis studies like \citealt{Breda_2018}. This method utilises the continuum morphology of the galaxy as the basis for computing radial statistics, offering a robust alternative to simple elliptical averaging. Isophotal annuli were constructed from the emission-line-free pseudo-continuum maps integrated over the rest-frame wavelength range $6390-6490$\AA, chosen to minimise the nebular contamination and to trace the stellar light distribution. From these maps, we defined a series of logarithmically equidistant isophotal zones (visible in Fig.~\ref{fig:isan}), each associated with a mean photometric radius (circularised).
The consecutive isophotal levels differ by a fixed step in surface brightness. This logarithmic spacing ensures sampling of the bright central regions and faint outskirts, while maintaining approximately uniform S/N across the annuli.

For each derived property (e.g. stellar mass surface density, mean stellar age), we computed the averaged value within each \textsc{isan} zone by applying the isophotal masks to the corresponding property map. This allowed us to produce high-fidelity 1D radial profiles for each quantity while accounting for the irregular morphology and inclination of the galaxies.
The advantage of the \textsc{isan} method lies in its preservation of structural integrity across complex galaxy morphologies, enabling consistent measurement of radial gradients and variations in physical properties both in central regions and outskirts. The derived profiles form the basis for subsequent analysis of stellar population gradients and galaxy assembly trends.

Because the radial extent sampled by \textsc{isan} zones varies across the galaxy sample, the number of galaxies contributing to the stacked radial profiles (Sections \ref{sec:ssd}, \ref{sec:ages},\ref{sec:met}, \ref{sec:ew} ) decreases at large radii. To minimise biases from incomplete radial coverage, averaged profiles for the intermediate-mass galaxies ($10 \leq \log(M_\star/M_\odot) < 11$) are limited to $\rm R/R_{eff}=3.0$, while the high-mass bin ($11 \leq \log(M_\star/M_\odot) \leq 11.5$) is extended to $\rm R/R_{eff}=3.25$, where sufficient radial coverage is still available. In the stacked profiles, dashed segments indicate radial bins where fewer than 50\% of galaxies contribute to the average.

\section{Results}
\label{results}
\subsection{Stellar surface density profiles}
\label{sec:ssd}
\begin{figure}[h]
    \centering
    \includegraphics[scale = 0.35]{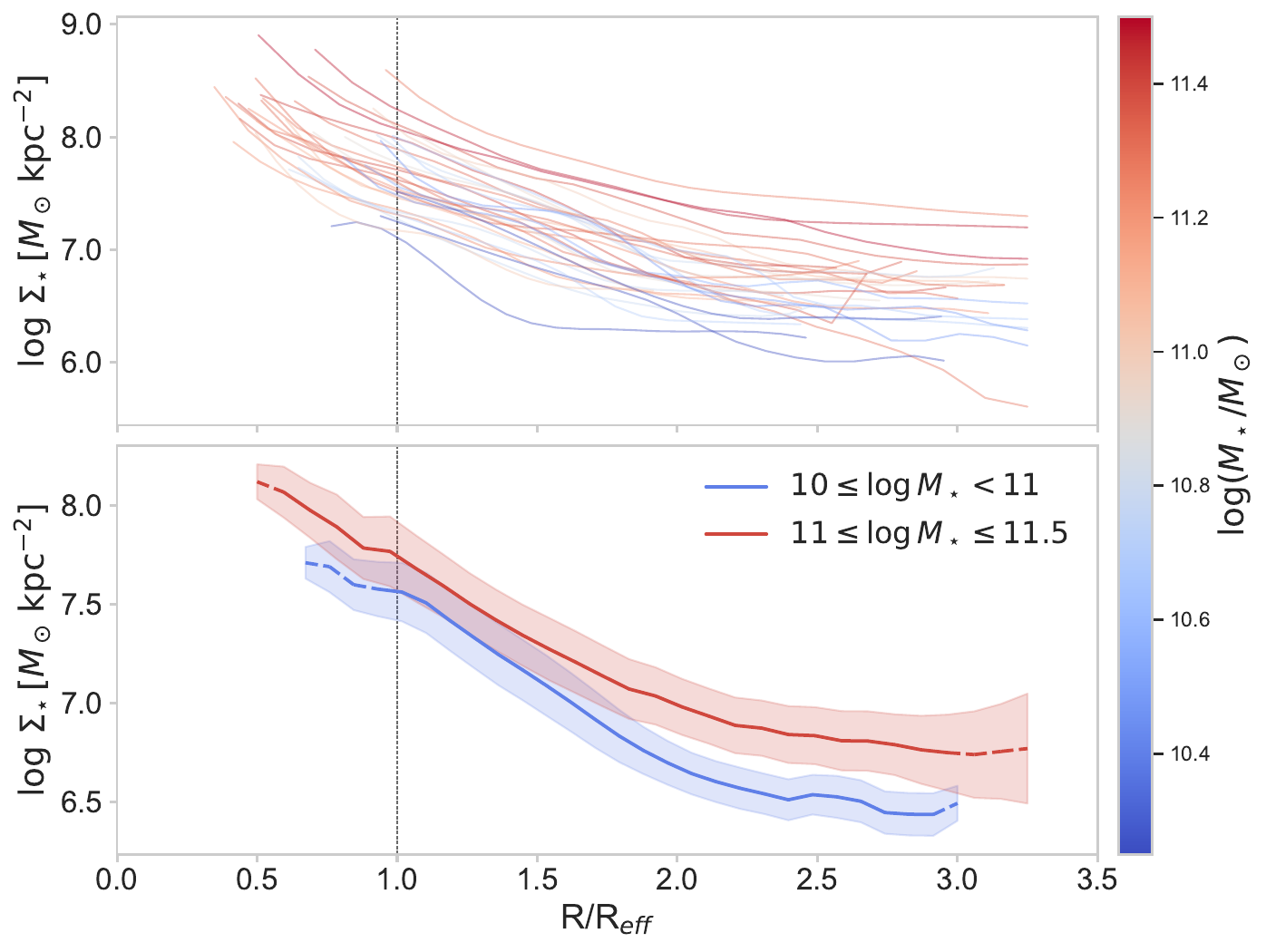}
    \caption{Radial profiles of stellar surface density, plotted as a function of radius normalised by the effective radius ($\rm R/R_{\mathrm{eff}}$) for all the galaxies in our sample, individually. The bottom panels show the averaged profiles interpolated for intermediate ($10 \leq \log(M_{\star}/M_{\odot}) < 11$) and high ($11 \leq \log(M_{\star}/M_{\odot}) \leq 11.5$) mass bins, in blue and red, respectively. Shaded regions correspond to 1 $\sigma$ of the average profiles. Solid lines indicate radial bins where at least 50\% of the galaxies contribute to the average profile, while dashed lines denote regions with lower radial coverage. The intermediate and high mass bins extend to $\rm R/R_{\mathrm{eff}} = 3.0$ and $3.25$, respectively. }
    \label{fig:mass_profiles}
\end{figure}
We begin by presenting the radial profiles of the stellar population properties to gain an insight into how these properties vary radially within the galaxies. In Fig.~\ref{fig:mass_profiles}, we show the radial profiles of stellar surface density ($\Sigma_\star$) for our sample of star-forming galaxies at $z \approx 0.3$ from the MAGPI survey (Section \ref{sec::magpi}). These profiles are constructed as a function of galactocentric distance, normalised by the effective radius ($\rm R_\mathrm{eff}$).
As expected, $\Sigma_\star$ exhibits a systematic decline with increasing radius, reflecting the centrally concentrated structure of disk galaxies. Such estimates serve as robust tracers of the spatial distribution of the stellar content across different galaxy regions.

Our galaxies show rapidly declining inner profiles and a pronounced flattening in the outskirts, beyond $\sim 2\: \rm R_{eff}$, which could be due to more extended SF activity or radial mixing. High-resolution work using HST and JWST resolved SED fitting \citep{Abdurrouf2023} also finds broadly negative $\Sigma_\star(r)$ gradients (normalised by half-mass radius) at $0.3< z<0.8$, with hints of shallower outer slopes at earlier times and higher radii, consistent with our results. Cosmological simulations also predict related behaviours \citep{Tacchella_2016a, Rodriguez-Gomez_2016, Pillepich_2018}.

However, it is important to note that, due to spatial resolution limits, the innermost parts ($<0.5\,R_{\rm eff}$) of our galaxies are not fully resolved, as demonstrated in Fig.~\ref{fig:isan}. Our inner slopes should therefore be considered conservative, PSF-smoothed values.

\subsection{Mass weighted and luminosity weighted quantities}
\subsubsection{Ages}
\label{sec:ages}
\begin{figure*}
    \centering
    \includegraphics[scale = 0.4]{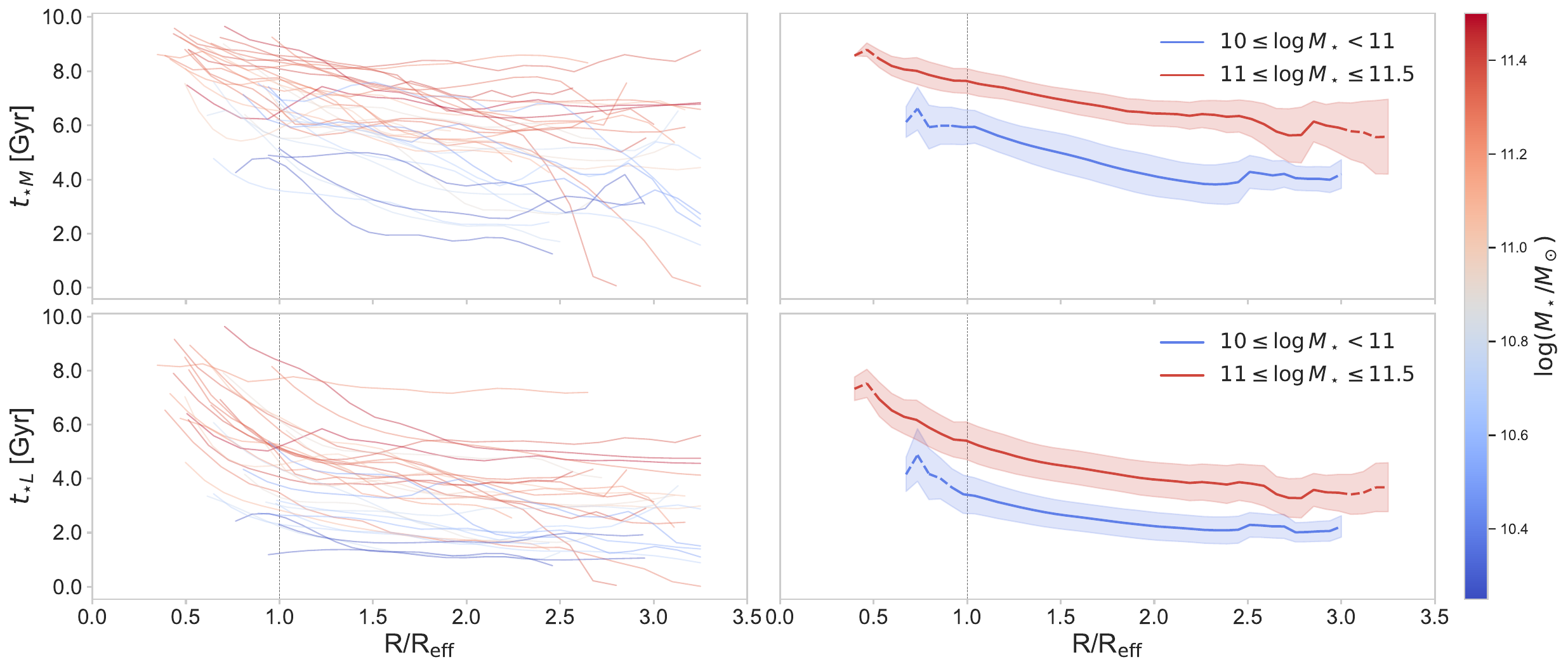}
\caption{Radial profiles of stellar ages using mass-weighted (top) and luminosity-weighted (bottom) averages, plotted as a function of radius normalised to $R_\mathrm{eff}$. Right panels show the averaged profiles interpolated for intermediate ($10 \leq \log(M_{\star}/M_{\odot}) < 11$) and high ($11\leq\log(M_{\star}/M_{\odot})\leq 11.5$) mass bins, in blue and red, respectively. Shaded regions correspond to 1 $\sigma$ standard deviation of the average profiles. Solid lines indicate radial bins where at least 50\% of the galaxies contribute to the average profile, while dashed lines denote regions with lower radial coverage. The intermediate and high mass bins extend to $\rm R/R_{\mathrm{eff}} = 3.0$ and $3.25$, respectively. }
\label{fig:t_radial}
\end{figure*}
Mass-weighted ages represent the formation epoch of stellar mass and trace the cumulative buildup of stellar mass, with older stars dominating. Luminosity-weighted ages, on the other hand, are more sensitive to the contribution of younger stellar populations due to their higher luminosities and thus capture more recent or ongoing SF activity. Including both types of profiles provides a more comprehensive view.

In Fig.~\ref{fig:t_radial} we present the radial profiles of stellar ages for our galaxy sample with mass-weighted ($t_{\star \rm M}$) and luminosity-weighted ($t_{\star \rm L}$) quantities, while a detailed quantitative comparison between weighted ages for every \textsc{isan} (corresponding to different $R_{\rm eff}$) is discussed in Appendix \ref{sec::radial}, Fig.~\ref{fig:tz_bd_compare}. The average profiles of the sample are shown in the right panels, interpolated for intermediate ($10 \leq \log(M_{\star}/M_{\odot}) < 11$) and high ($11\leq\log(M_{\star}/M_{\odot})\leq 11.5$) mass bins, in blue and red, respectively. We observe clear negative radial gradients in both age metrics, with stellar ages systematically decreasing from the galaxy centres to their outskirts.

The curve's steepness is not uniform across the different regions. The inner $0.5-1\,R_{\rm eff}$ parts (Fig.~\ref{fig:t_radial}, left panel) generally exhibit a steeper slope. Age gradients tend to flatten, with only mild negative slopes, as the radial profiles decline consistently. 

In the range $1-2\,R_{\rm eff}$, the profiles mark a transition region where the SFHs are more diverse. For the average mass-weighted age profiles, high-mass galaxies (red curves) exhibit a relatively shallow radial decline, with $t_{\star \rm M}$ decreasing from $\sim9$ Gyr in the inner regions to $\sim7.5$ Gyr at $1\,R_{\rm eff}$, and further down to $\sim6.5$ Gyr by $2\,R_{\rm eff}$. In contrast, intermediate-mass galaxies (blue curves) display a different radial behaviour. As already suggested by the individual profiles in the left panel, where a mass dependence is visible, the averaged profile is flatter in the inner regions, only slightly decreasing from $6.5-7$ Gyr to $6$ Gyr within $1\,R_{\rm eff}$. Beyond this radius, the profile shows a more pronounced decline, reaching $\sim5$ Gyr at $2\,R_{\rm eff}$.

The luminosity-weighted age profiles show similar radial trends across both mass bins, with the high and intermediate-mass averages offset by $1-2$ Gyr at all radii. In the inner regions, both profiles show declines of $t_{\star \rm L}$ decreasing from $7.5$ Gyr to $5.5$ Gyr for the high-mass bin, and from $5$ Gyr to $3$ Gyr for the intermediate-mass bin within $1\,R_{\rm eff}$. Out to $2\,R_{\rm eff}$, the gradients become noticeably shallower. Over this radial range, the total decrease amounts to only $\sim1.25$ Gyr for the high-mass galaxies and $\sim1$ Gyr for the intermediate-mass systems, indicating a significant flattening of the luminosity-weighted age profiles.

At larger radii ($R \geq 2\,R_{\rm eff}$), both averaged $t_{\star \rm M}$ and $t_{\star \rm L}$ profiles tend to flatten within the uncertainties. By $R = 3 \,R_{\rm eff}$ and $R=3.25 \,R_{\rm eff}$, the intermediate-mass and high-mass bins converge to values of $t_{\star \rm M} \sim 6 $ Gyr and $t_{\star \rm M} \sim 5.5$ Gyr, respectively, suggesting that the outermost stellar populations are systematically younger and formed over more extended timescales. For $t_{\star \rm L}$, no significant variations are observed as the intermediate-mass profile remains essentially constant, while the high-mass profile shows only a marginal additional decrease of $\sim0.25$ Gyr.

Importantly, the shapes of the radial age profiles show a clear dependence on stellar mass. The averaged profiles for the higher mass bin (red curves) are systematically older at all radii than the intermediate-mass bin (blue curves). This behaviour implies that more massive systems assembled their central stellar mass earlier and experienced less late-time SF in their outskirts, while lower-mass systems continued to grow their outer parts at later epochs, yielding larger differences between mass- and light-weighted ages in the outer parts. 

An important aspect of our analysis is the comparison between the two age weightings. At large radii ($R > 2\,R_{\rm eff}$), average $t_{\star}$ profiles agree, however, individual profiles (Fig.~\ref{fig:t_radial} left panel), especially for $t_{\star \rm M}$, show more diversity in the outer slopes, which could be taken as evidence for continued or recent low-level SF in the outer disk. $t_{\star \rm M}$ continues to decline slightly in this region, indicating that an older stellar component is still present even in the outskirts, but not as significantly. We caution that the number of spatial bins contributing to the stacked averages declines rapidly beyond $\sim2.5\,R_{\rm eff}$, particularly for the lower-mass bin (see Fig.~\ref{fig:isan} for annuli distribution as a function of $R_{\rm eff}$). As a result, measurements at $R\geq2.5\,R_{\rm eff}$ are driven by a smaller and potentially biased subset of the sample (e.g. more extended or more face-on systems). 

\subsubsection{Metallicities}
\label{sec:met}
\begin{figure*} 
    \centering
    \includegraphics[scale = 0.4]{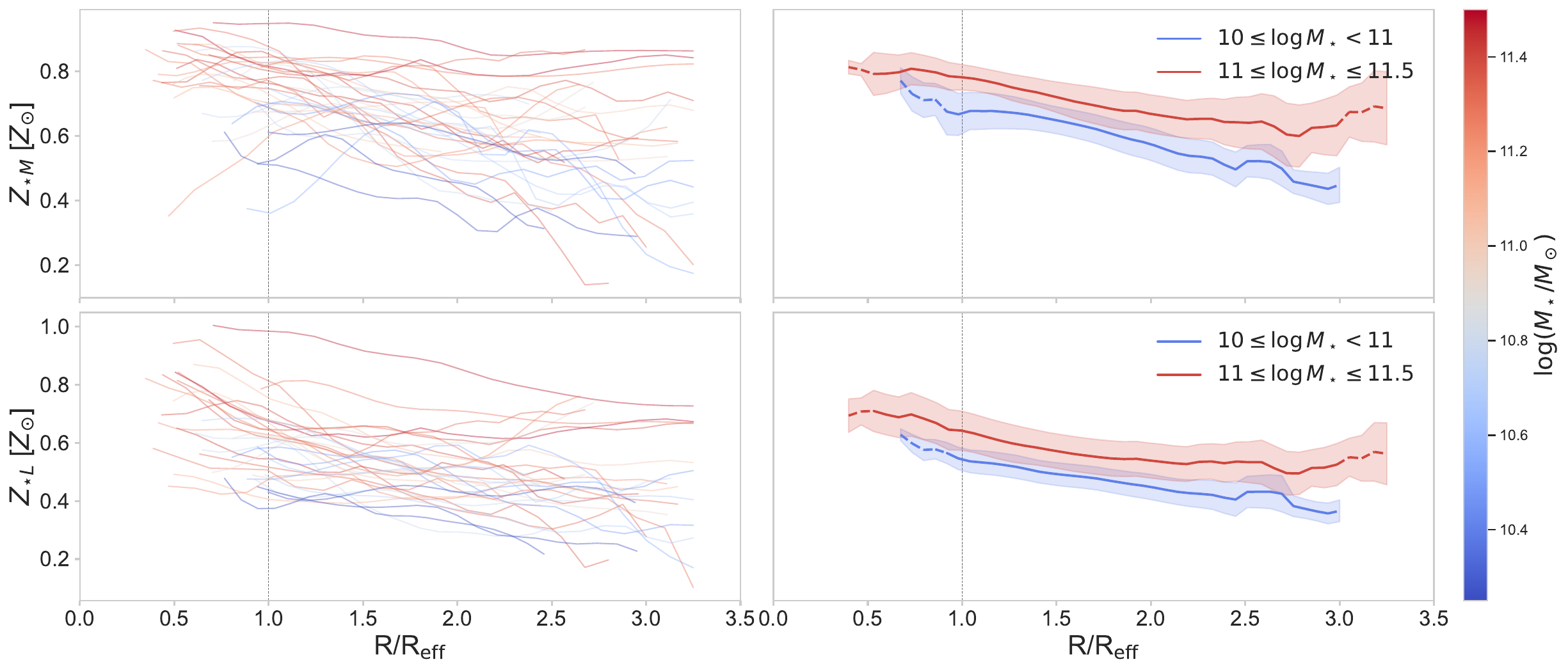}
    \caption{Radial profiles of stellar metallicities using mass-weighted (top) and luminosity-weighted (bottom) averages, plotted as a function of radius normalised to $R_\mathrm{eff}$. Right panels show the averaged profiles interpolated for intermediate ($10 \leq \log(M_{\star}/M_{\odot}) < 11$) and high ($11 \leq \log(M_{\star}/M_{\odot}) \leq 11.5$) mass bins, in blue and red, respectively. Shaded regions correspond to 1 $\sigma$ standard deviation of the average profiles. Solid lines indicate radial bins where at least 50\% of the galaxies contribute to the average profile, while dashed lines denote regions with lower radial coverage. The intermediate and high mass bins extend to $\rm R/R_{\mathrm{eff}} = 3.0$ and $3.25$, respectively. }
    \label{fig:z_radial}
\end{figure*}
In a similar way to age profiles, the distinction between mass and luminosity weighted metallicities reflects the different stellar populations that dominate each weighting. $Z_{\star M}$ traces the cumulative chemical enrichment of the stellar component and is dominated by older, long-lived stars that preserve the metallicity of the interstellar medium at the time of their formation. Therefore, it provides a record of the integrated past enrichment history and the efficiency of metal retention over cosmic time. In contrast, $Z_{\star L}$ is more strongly influenced by the younger, brighter stellar populations, which formed from gas recently enriched by inflows, feedback, or mixing processes.
Fig.~\ref{fig:z_radial} presents the radial distributions of stellar metallicities for our galaxy sample, derived using both mass-weighted ($Z_{\star M}$) and luminosity-weighted ($Z_{\star L}$) quantities. The panels on the left report the individual metallicity profiles for each galaxy in our sample, colour-coded for total galaxy stellar mass, while the right panels show the interpolated averaged curved for intermediate ($10 \leq \log(M_{\star}/M_{\odot}) < 11$) and high ($11\leq\log(M_{\star}/M_{\odot})\leq 11.5$) mass bins, in blue and red, respectively.

Across the full sample, we detect clear negative radial gradients in both stellar metallicity profiles. Within the inner regions ($R \leq 1\,R_{\rm eff}$), the average mass-weighted metallicity decreases by $0.05$ dex and $0.10$ dex for the high-mass and intermediate-mass regimes, respectively. Although luminosity-weighted metallicity profiles of individual galaxies, particularly in the upper high-mass regime, can appear steeper in the inner regions, the averaged profiles exhibit a decline comparable to that of the mass-weighted metallicities, of the order of $\sim0.05-0.10$ dex.

Beyond $1\,R_{\rm eff}$, negative gradients become more pronounced in all four averaged profiles, while still showing a similar behaviour between the two weighting schemes. Specifically, the mass-weighted metallicity $Z_{\star M}$ decreases from $0.78$ to $0.66\,Z_{\odot}$ in the high-mass regime and from $0.67$ to $0.57\,Z_{\odot}$ in the intermediate-mass regime. Similarly, the luminosity-weighted metallicity, $Z_{\star L}$, declines from $0.64$ to $0.53\,Z_{\odot}$ and from $0.54$ to $0.45\,Z_{\odot}$ for the high- and intermediate-mass bins, respectively. For both $Z_{\star M}$ and $Z_{\star L}$, the outer slopes continue to decrease steadily, reaching values of $0.45\,Z_{\odot}$ and $0.36\,Z_{\odot}$ at a radius of up to $3\,R_{\rm eff}$.

Beyond $2.75\,R_{\rm eff}$, the averaged metallicity profiles, particularly for the higher-mass galaxies, become increasingly uncertain, as indicated by the rapidly rising scatter. In this radial range, individual galaxies display a wide variety of behaviours, with some exhibiting nearly flat metallicity profiles and others showing sharp drops of up to $\sim0.2$ dex. Consequently, the mean trends at these radii are strongly affected by galaxy-to-galaxy variations and limited radial coverage, and should therefore be interpreted with caution. Overall, higher-mass galaxies exhibit systematically higher metallicities at all radii and shallower radial gradients, in agreement with the established stellar mass-stellar metallicity relation \citep{Dom_nguez_G_mez_2023, looser2024, baker2024}.

At all radii, the mass-weighted metallicity exceeds the luminosity-weighted metallicity by $0.05-0.10$ dex. This offset reflects the increasing contribution of younger, lower-metallicity stellar populations at large galactocentric distances, which dominate the emitted light but contribute less to the total stellar mass. The observed mass dependence supports a scenario in which massive galaxies completed their chemical enrichment earlier and more efficiently, whereas lower-mass systems continue to build up metals in their outskirts at $z \sim 0.3$, consistent with inside-out growth and downsizing-driven galaxy evolution.

We note that a detailed quantitative comparison between the mass-weighted and luminosity-weighted metallicities for every \textsc{isan} (corresponding to different $R_{\rm eff}$) is discussed in Appendix \ref{sec::radial} (Fig.~\ref{fig:tz_bd_compare}). Fig.~\ref{fig:tz_mass_dep} also shows direct trends between stellar metallicities, ages, and corresponding \textsc{isan} masses.

\subsection{H$\alpha$ equivalent widths profiles} 
\label{sec:ew}
\begin{figure} 
    \centering
    \includegraphics[scale = 0.375]{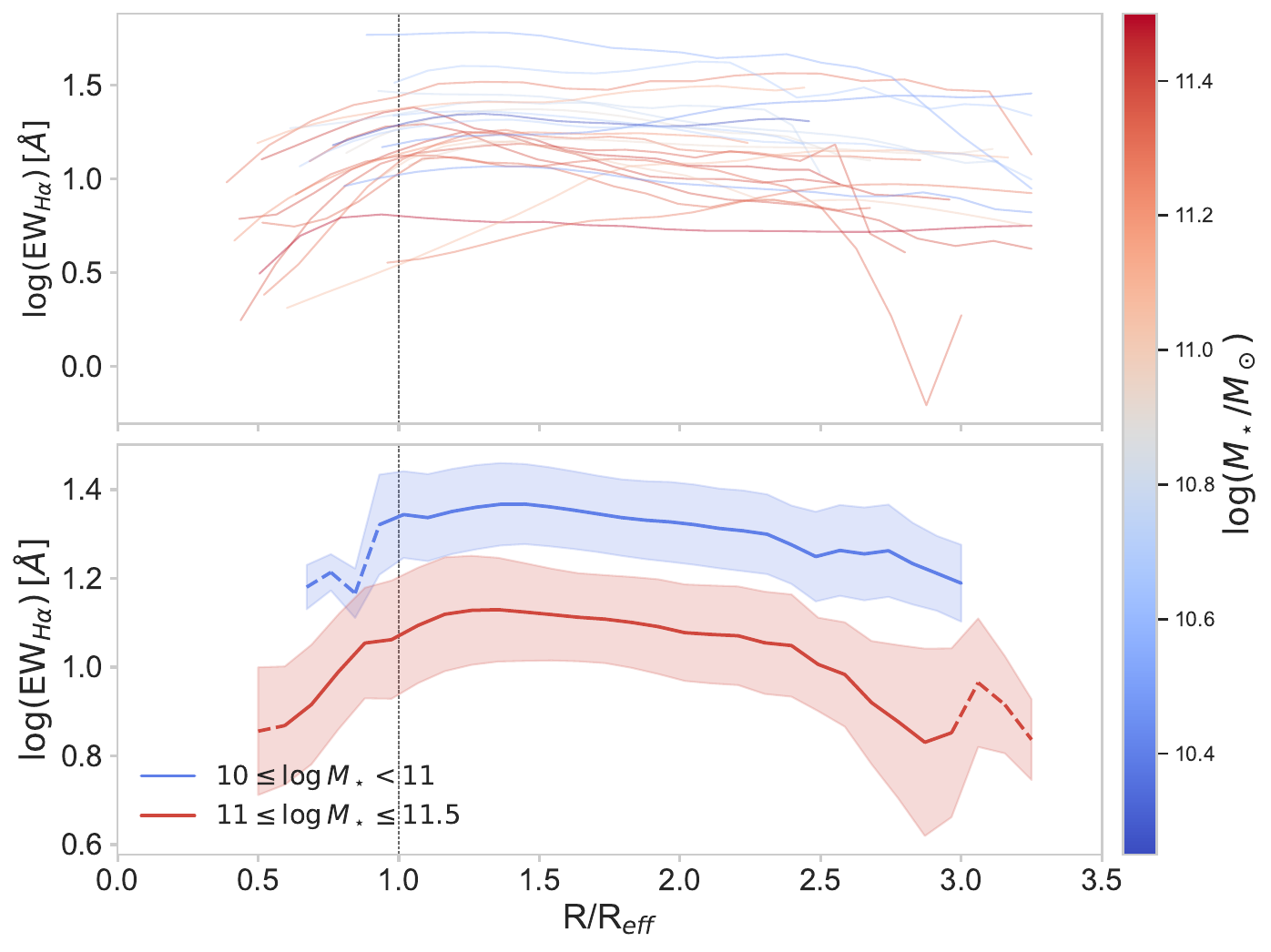}
    \caption{Radial profiles of H$\alpha$ equivalent width, plotted as a function of radius normalised to $R_\mathrm{eff}$. In the top panel are the radial profiles of H$\alpha$ equivalent width for galaxies in our sample that do not exhibit strong AGN features in the innermost bins. The bottom panel shows the averaged profiles interpolated for intermediate ($10 \leq \log(M_{\star}/M_{\odot}) < 11$) and high ($11 \leq \log(M_{\star}/M_{\odot}) \leq 11.5$) mass bins, in blue and red, respectively. Shaded regions correspond to 1 $\sigma$ standard deviation of the average profiles. We note that the y axis ranges differ between the upper and lower panels. Solid lines indicate radial bins where at least 50\% of the galaxies contribute to the average profile, while dashed lines denote regions with lower radial coverage. The intermediate and high mass bins extend to $\rm R/R_{\mathrm{eff}} = 3.0$ and $3.25$, respectively.}
    \label{fig:EW_radial}
\end{figure}
The equivalent width of H$\alpha$ (EW$_{H\alpha}$) measures the H$\alpha$ line flux relative to the underlying stellar continuum and therefore provides a tracer of the specific star formation rate (sSFR) and of the dominant ionisation mechanisms across galaxy disks. While the absolute H$\alpha$ flux distribution typically peaks towards the central regions and traces the instantaneous star formation rate, EW$_{H\alpha}$ is more sensitive to the contrast between current and past SF, provided it is not significantly contaminated by active galactic nucleus (AGN) activity. High EW$_{H\alpha}$ values thus indicate regions where recent SF dominates the luminosity, whereas low values may arise from quenching, AGN ionisation, or emission from evolved stellar populations such as post-AGB stars \citep{Belfiore2017}.

The top panel of Fig.~\ref{fig:EW_radial} shows the radial EW$_{H\alpha}$ profiles for our sample, after excluding systems dominated by strong AGN emission (7 galaxies), identified using BPT diagnostic criteria in their innermost bins. The average profiles exhibit low central EW$_{H\alpha}$ values, typically below $30\,\AA$ ($\log \mathrm{EW}_{H\alpha}\approx1.5$), followed by a steep rise out to $1\,R_{\rm eff}$. Beyond this radius, the profiles flatten, up to $2.5\,R_{\rm eff}$, with a mild decline at larger radii. This overall behaviour is consistent with that reported by \cite{Kalinova_2021} for non-active and weakly active AGN galaxies and reflects centrally suppressed SF combined with enhanced or ongoing activity in the outskirts.

The bottom panel of Fig.~\ref{fig:EW_radial} reveals a clear stellar mass dependence in the EW$_{H\alpha}$ radial profiles. Both intermediate-mass galaxies ($10<\log M_\star/M_\odot<11$) and more massive systems ($\log M_\star/M_\odot>11$) show a steep increase in EW$_{H\alpha}$ within the inner $1\,R_{\rm eff}$. In this region, intermediate-mass galaxies reach $\log(\mathrm{EW}_{H\alpha})\sim 1.33$, while more massive galaxies attain lower values of $\log(\mathrm{EW}_{H\alpha})\sim 1.10$. At larger radii, the profiles flatten significantly up to $2.5\,R_{\rm eff}$, where the slopes change by only $0.1$ dex, with typical EW$_{H\alpha}$ values of $1.25$ for lower-mass and $1.0$ for higher-mass systems.

Lower-mass galaxies therefore exhibit higher EW$_{H\alpha}$ values at all radii, indicating that their specific star formation remains elevated across the outskirts. In contrast, more massive galaxies exhibit lower overall EW$_{H\alpha}$ levels, consistent with increasing dominance of old stellar populations. In these systems, EW$_{H\alpha}$ may be suppressed not only by reduced ongoing SF but also by enhanced stellar continuum emission and H$\alpha$ absorption associated with older stellar populations.

The decline observed in the outermost radial bins ($R > 2.5\,R_{\rm eff}$) should be interpreted with caution, as it is likely influenced by decreasing surface brightness, reduced spatial coverage, and the smaller number of spaxels contributing at large radii. These effects can increase uncertainties and bias the mean EW$_{H\alpha}$ measurements in the outer regions. For completeness, Appendix \ref{sec::radial} and Fig.~\ref{fig:ew_vs_stars} present the EW$_{H\alpha}$ values measured for all individual \textsc{isan} zones, in connection with the other properties extracted from the radial profiles, including stellar mass,  and mass-weighted and luminosity-weighted stellar ages and metallicities, separately for the inner and outer regions.

\subsection{Inner regions versus outer regions}
\begin{figure} 
    \centering
    \includegraphics[scale = 0.45]{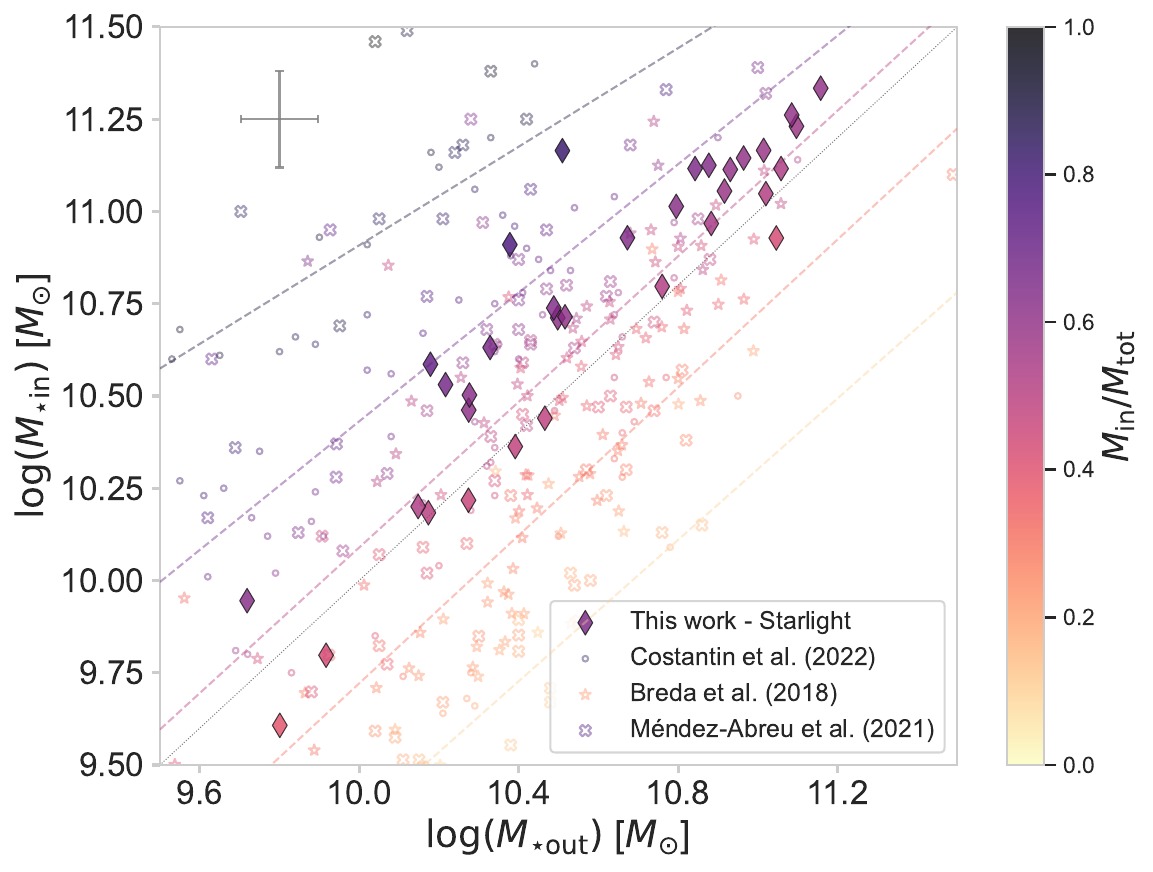}
    \caption{Inner region versus outer region stellar mass for our galaxy sample, compared with previous spectro-photometric decomposition studies: \cite{Costantin2022} (circles), \cite{Breda_2018} (stars), and \cite{MendezAbreu2021} (crosses). Dashed lines represent the best-fitting relations between bulge (and inner regions) and disk (outer regions) mass across the different mass ratios for the data samples mentioned above, colour-coded according to the colour bar. The dotted grey line indicates the 1:1 bisector.}
    \label{fig:mass_bd}
\end{figure}
Throughout this study, we refer to physical quantities measured within the inner and outer regions of our galaxy sample. These regions are defined using a purely radial criterion based on the effective radius, such that the inner region comprises all spaxels within $1\,R_{\rm eff}$, while the outer region includes all spaxels located at radii larger than $1\,R_{\rm eff}$. All stellar population properties, including stellar masses and mass and luminosity-weighted parameters, are extracted by averaging over the full bin extent of the corresponding region, unless stated otherwise.

Fig.~\ref{fig:mass_bd} shows the relation between the stellar masses of the inner and outer regions in our sample, compared with results from recent spectro-photometric and dynamical decomposition studies. We find a strong positive correlation between inner-region mass ($M_{\star \,\rm in}$) and outer-region mass ($M_{\star \,\rm out}$), spanning $ 9.72\leq \log(M_{\star\, \rm out}/M_\odot) \leq 11.15$ and $ 9.60\leq \log(M_{\star\, \rm in}/M_\odot) \leq 11.30$. 

The distribution broadly agrees with previous IFS-based studies, particularly \cite{Breda_2023} and \cite{MendezAbreu2021}, which derived stellar masses for distinct structural components. Galaxies with lower outer-region masses are typically dominated by their extended disks, while the inner regions become more prominent as total mass increases. For instance, systems around $\log(M_{\star \, \rm out}/M_\odot) \sim 10.3$ host inner regions with $\log(M_{\star\, \rm in}/M_\odot) \sim 10.1-10.5$ ($M_{\star \, \rm in}/M_{\rm tot} \sim 0.4-0.5$). 

As total stellar mass increases, the contribution of the inner regions becomes progressively more significant, with the most massive systems in our sample reaching $M_{\star \, \rm in}/M_{\rm tot} > 0.5$, indicative of a transition towards centrally dominated morphologies. Owing to the nature of our decomposition, we expect a mild overestimation of the inner-region mass and a corresponding underestimation of the outer-region mass; these values should therefore be regarded as upper and lower limits, respectively. This would also affect the mass ratios (shown in different colours in the plot), as the majority of our galaxies exhibit high inner-to-total mass ratios. The dashed coloured lines in Fig.~\ref{fig:mass_bd} indicate linear fits corresponding to fixed inner-to-total stellar mass ratios ($M_{\star \, \rm in}/M_{\rm tot}$), in steps of 0.2, from 0 to 1, obtained from the combined dataset including both this work and the literature samples. The colour coding matches that of the data points colour bar. Our measurements closely follow the same colour sequence as the other comparison samples, despite representing upper and lower limits on the inner and outer stellar masses.

Unlike \cite{Breda_2018}, we do not find a clear low-mass reversal in the $M_{\star , \rm in}-M_{\star , \rm out}$ relation, partly due to the limited number of low-mass systems. At the high-mass end, our results converge with those of \cite{Breda_2018} and \cite{MendezAbreu2021}, who found massive spirals to host old, metal-rich, high-Sérsic-index central structures consistent with classical bulges. Compared to \cite{Costantin2022}, our inner-to-total mass fractions are shifted towards lower values and higher outer masses, with a larger fraction of galaxies at high $M_{\star , \rm in}/M_{\rm tot}$, likely reflecting differences in sample selection and decomposition methods rather than population differences.

\subsubsection{Inner regions: Ages and metallicities}
\label{sec:inner}
\begin{figure*} 
    \centering
    \includegraphics[scale = 0.365]{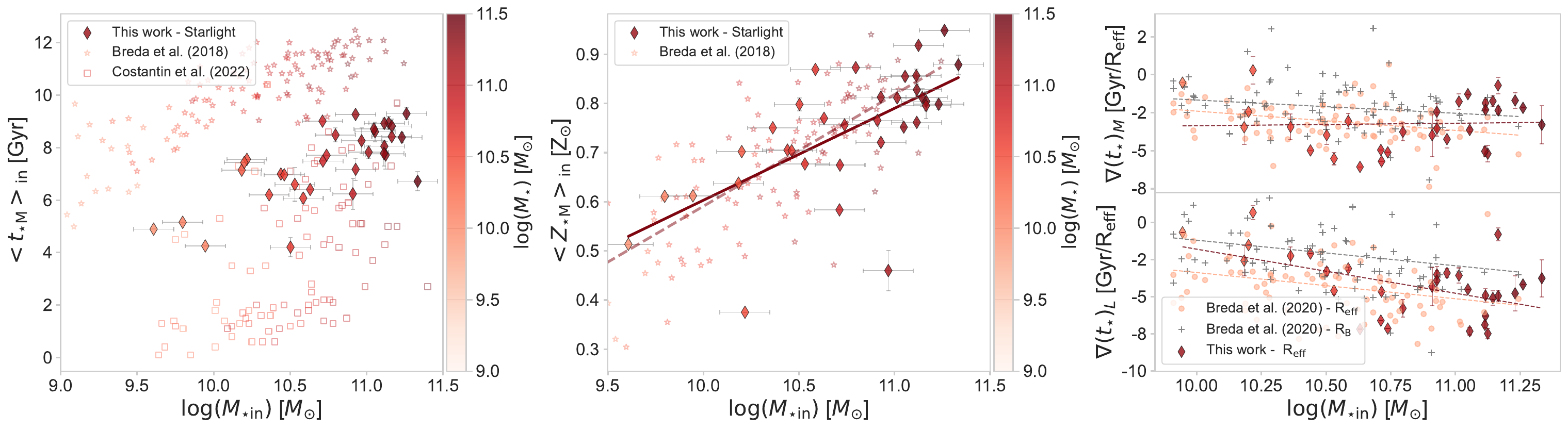}
    \caption{Stellar mass-weighted ages (left), metallicities (centre), and age gradients (right) as a function of stellar mass for inner ($R \leq R_{\mathrm{eff}}$) regions. In the left and central panels, results from this work with \texttt{Starlight} (diamonds) are shown alongside those from \cite{Costantin2022} (squares) at $0.14\leq z \leq 1$ and \cite{Breda_2018} (stars) at $z =0$. 
    The mass and light weighted stellar age gradients measurements $\nabla(t_{\star})_M$ and $\nabla(t_{\star})_L$ are in the upper and bottom panel, respectively. The results from this work, in \texttt{Starlight}, are directly compared with the results from \cite{Breda_2020} in the local Universe. In particular, circles refer to the gradients calculated with $R_{\rm eff}$, while the grey crosses refer to the gradients derived within the photometrically decomposed bulge, with $R_{\rm B}$ being the bulge radius. For the inner regions of this work, the gradients were calculated only for the galaxies with at least two isophotal annulus measurements within 1 $R_{\rm eff}$. The dashed lines represent the linear fits for each data sample. All data points are colour-coded by total galaxy stellar mass, as indicated by the colour bars, with red indicating the inner and bulge regions.}
    \label{fig:age_z_bd_in}
\end{figure*}
Fig.~\ref{fig:age_z_bd_in} presents the relation between mass-weighted stellar ages and metallicities as a function of stellar mass for the inner regions of our galaxies, compared with measurements from \cite{Breda_2018} and \cite{Costantin2022}. These studies use accurate spectral-photometric decomposition methods to distinguish galaxy bulges and disks. They provide an insight into the SFHs of distinct structural components, tracing their relative growth channels and assembly timescales. The inner regions exhibit a tight age-mass relation. For galaxies with $\log(M_{\star \rm in}/M_\odot) > 10.5$, inner-region ages typically lie in the range $t_{\star \, \rm in}\approx 6-10$ Gyr. Both normalisation and scatter of this relation closely resemble those observed in local samples, once differences in look-back time ($\sim 3.5$ Gyr) are accounted for.

As for the metallicities, they span $\langle Z_{\star\,\rm in}\rangle_M = 0.38-0.95\,Z_{\odot}$ over $\log(M_{\star\, \rm in}/M_{\odot}) \approx 9.6-11.3$. For massive inner regions ($\log(M_{\star\, \rm in}/M_\odot)\geq10.75$), metallicities cluster tightly around $0.8-0.95\,Z_\odot$ with a small dispersion ($\lesssim0.1$ dex), indicating uniformly high enrichment in the central parts of massive galaxies. We obtain
$\langle Z_{\star}\rangle_M/Z_\odot = 0.17\cdot\log(M_{\star}/M_\odot) - 1.19$, steeper than what \cite{Breda_2018} found for bulges ($\langle Z_{\star}\rangle_M/Z_\odot = 0.20\cdot\log(M_{\star}/M_\odot) - 1.57$). Thus, the slopes indicate that the mass dependence of chemical enrichment is very similar between the two samples. 

The rightmost panel of Fig.~\ref{fig:age_z_bd_in} compares the mass and light-weighted stellar age gradients derived in this work with those from \cite{Breda_2020} in the local Universe, using a galaxy sample from CALIFA. For the local galaxy sample, we show not only the gradients derived in inner regions within 1 $R_{\rm eff}$ (light pink), but also the ones estimated within the bulge radius, $R_{\rm B}$ (grey), using their photometric bulge-disk decomposition. In the inner mass-weighted regions (top panel), our gradients appear overall significantly flatter than locally (dashed red line fit). This difference is clearly reflected in the linear fits: while \cite{Breda_2020} report slopes of $m = - 0.87$ (within $R_{\rm B}$) and $m = -1.27$ (within $R_{\rm eff}$), our relation is consistent with a nearly flat trend, with $m = 0.17$ (with intercepts of $7$, $10.3$, $-5$, respectively). Despite this flat average behaviour, there is a significant scatter in the individual measurements, with $\nabla(t_{\star})_M$ spanning values from $0$ to $-6\,{\rm Gyr}/R_{\rm eff}$. This suggests that, even if the central regions of our galaxies are undergoing quenching or are already quenched, they still experience residual activity and chemical enrichment. 

Such ongoing or recently ceased SF may smooth the stellar age distribution, producing less negative slopes. Part of this difference may also arise from sample demographics and spatial resolution, as the smaller number of radial points in the innermost regions introduces larger uncertainties in the gradient estimates.
The inner luminosity-weighted gradients (bottom panel) show reasonable agreement with those of local galaxies. We recover a steep dependence on inner region stellar mass, with a slope of $m = -2.95$, steeper than the values reported by \cite{Breda_2020} of $m = -1.76$ within $R_{\rm eff}$ and $m = -1.72$ within $R_{\rm B}$ (with intercepts of $27.8$, $14.2$, and $15.9$, respectively). The gradients span values from $\nabla(t_{\star})_L \approx -0.5$ up to $ -6\,{\rm Gyr}/R_{\rm eff}$ over the explored mass range, supporting the idea that these regions are dominated by evolved stellar populations whose light is only mildly affected by younger stars.

\subsubsection{Inner regions: Star formation}
\label{sec::ssf_in}
\begin{figure}
    \centering
    \includegraphics[scale = 0.5]{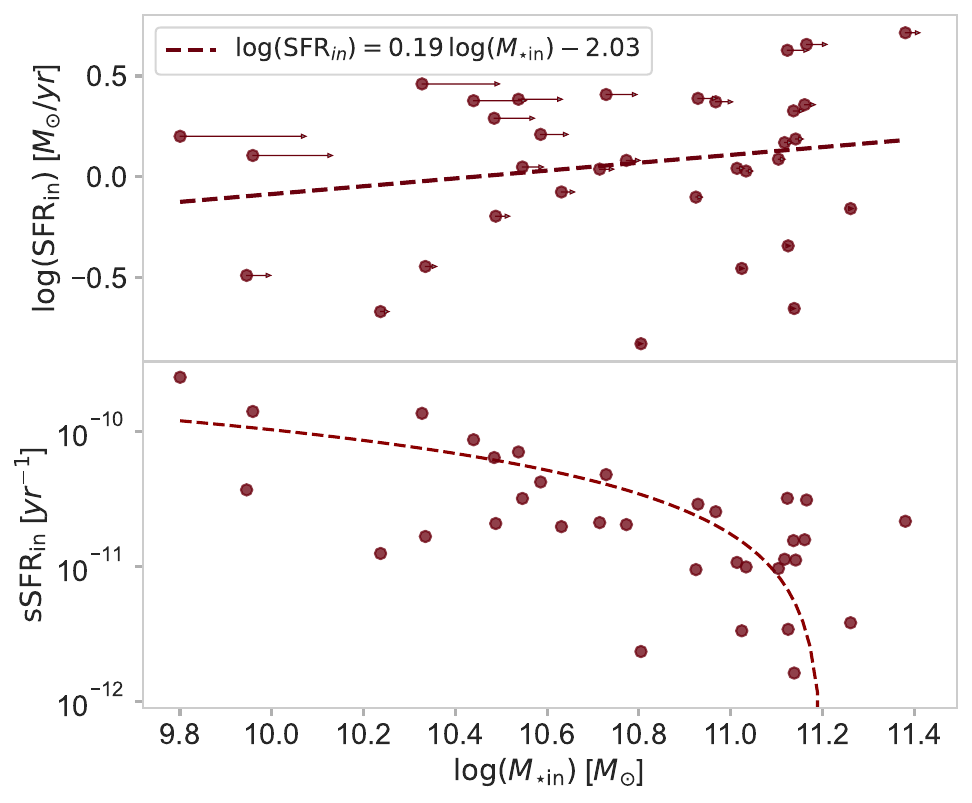}
    \caption{Top panel: Logarithmic SFR as a function of component stellar mass for inner regions. Arrows indicate the stellar mass that would be formed if the current SFR were constant between $z\sim0.3$ and $z=0$ with no mass loss, illustrating the potential impact of ongoing SF on component growth. Bottom panel: Specific star formation rate (sSFR) as a function of component stellar mass, fitted in a logarithmic scale. The properties presented are the result of this work with \texttt{Starlight}.}
    \label{fig:sfr_in}
\end{figure}
In Fig.~\ref{fig:sfr_in}, we examine the star-formation rates (SFRs) and specific star formation rates (sSFRs) of the inner regions in our sample as a function of their stellar mass.  Arrows indicate the stellar mass that each region would hypothetically accrete if its current mean SFR were sustained over the corresponding look-back time, neglecting mass loss and recycling. This representation provides an intuitive estimate of the potential contribution of ongoing SF to the present-day stellar-mass budget of the inner components.

The inner regions exhibit systematically lower SF activity across the full region stellar-mass range considered, with $\log(\mathrm{SFR}_{\rm in}) \lesssim 0.5~[M_\odot\,\mathrm{yr}^{-1}]$. Although the mean SFR increases with increasing stellar mass, the scatter is large, and the inferred potential accreted mass (arrows) tends to be larger at lower masses, reflecting both higher relative growth rates and longer effective timescales.

This behaviour is further emphasised by the sSFRs shown in the bottom panel of Fig.~\ref{fig:sfr_in}. The inner sSFR declines steeply with increasing stellar mass, with typical values of $\log(\mathrm{sSFR}_{\rm in}/\mathrm{yr}^{-1}) \sim -10.0$ at $\log(M_{\rm in}/M_\odot)\sim10$, decreasing to below $-11$ at the high-mass end. This trend indicates that the inner regions of massive galaxies are probably quenched and contribute negligibly to the current stellar-mass assembly. 

\subsubsection{Outer regions: Ages and metallicities}
\label{sec:outer}
\begin{figure*} 
    \centering
    \includegraphics[scale = 0.365]{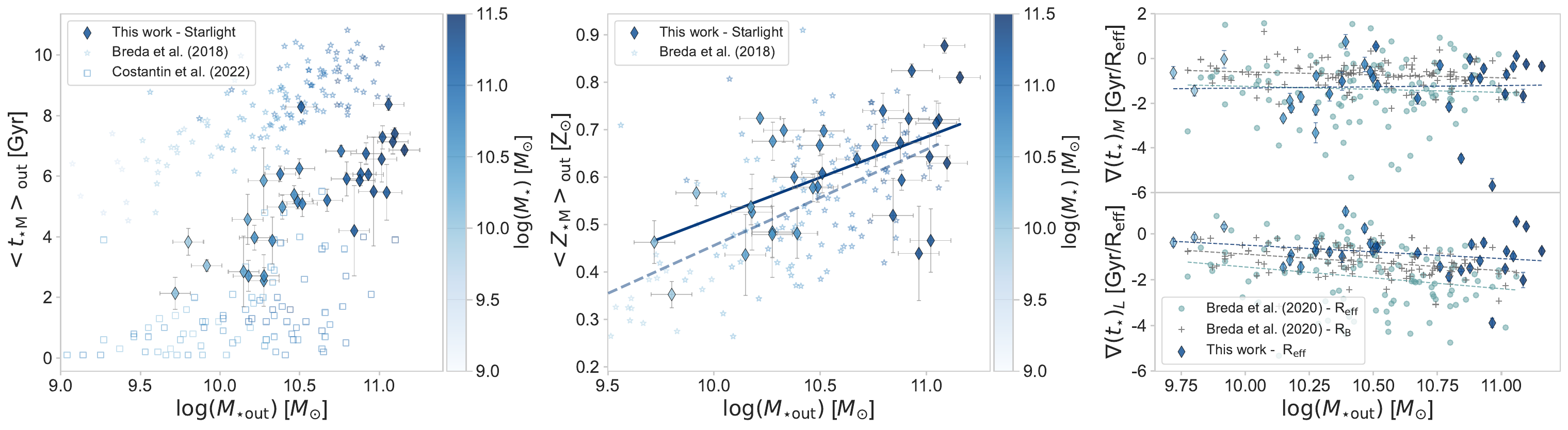}
    \caption{Stellar mass-weighted ages (left), metallicities (centre), and age gradients (right) as a function of stellar mass for outer ($R > R_{\mathrm{eff}}$) regions. In the left and central panels, results from this work with \texttt{Starlight} (diamonds) are shown alongside those from \cite{Costantin2022} (squares) at $0.14\leq z \leq 1$ and \cite{Breda_2018} (stars) at $z = 0$. 
    The mass and light weighted stellar age gradients measurements $\nabla(t_{\star})_M$ and $\nabla(t_{\star})_L$ are in the upper and bottom panel, respectively. The results from this work, in \texttt{Starlight}, are directly compared with the results from \cite{Breda_2020} in the local Universe. In particular, circles refer to the gradients calculated with $R_{\rm eff}$, while the grey crosses refer to the gradients derived outside the photometrically decomposed radius, with $R_{\rm B}$ being the bulge radius. The dashed lines represent the linear fits for each data sample. All data points are colour-coded as a function of total galaxy stellar mass, as reported in the colour bars, in blue for outer and disk regions.}
    \label{fig:age_z_bd_out}
\end{figure*}
Similar to Fig.~\ref{fig:age_z_bd_in}, Fig.~\ref{fig:age_z_bd_out} reports the relation between stellar mass-weighted ages and metallicities as a function of stellar mass for the outer ($R > R_{\mathrm{eff}}$) regions of our sample. Outer regions ages exhibit a positive correlation between $t_{\star\,\rm out}$ and $\log(M_{\star\, \rm out})$, with mass-weighted ages spanning from $2$ to $8$ Gyr across the sampled mass range, but are systematically younger than their inner counterparts, especially for lower mass systems. At low masses ($\log(M_{\star\, \rm out}/M_\odot)\leq10.5$), the age distribution is broad, with $t_{\star\, \rm out}\approx2-6$ Gyr. In contrast, more massive outer regions ($\log(M_{\star\, \rm out}/M_\odot) >10.5$) are systematically older, with typical ages up to $6$ and $8$ Gyr and a reduced scatter ($\lesssim1$ Gyr), suggesting earlier quenching of SF. 

The shape and slope of the outer-region age-mass relation are in good agreement with those reported by \cite{Breda_2018} for local galaxies and also with those of \cite{Costantin2022} once the look-back time of the individual galaxies is taken into account. At a fixed stellar mass, however, our galaxies are younger by $3-4$ Gyr. When interpreted in terms of cosmic time, this offset is consistent with the look-back time difference between our sample ($z\sim0.3$) and the present-day Universe, indicating predominantly passive evolution of the outer stellar populations since $z\sim0.3$, with little evidence for widespread rejuvenation or late secondary SF episodes in massive systems (see also Section \ref{sec::sfh}).

Metallicity-wise, the outer regions are systematically less metal-rich than the inner regions at a fixed stellar mass, spanning $\langle Z_{\star\, \rm out}\rangle_M \approx 0.35-0.87\,Z_\odot$ across $\log(M_{\star\, \rm out}/M_\odot)\sim9.7-11.15$. The distribution is broad at both the lower- and higher-mass ends, with scatter reaching $0.4$ dex in the more massive regions. Consistent with heterogeneous enrichment histories and continued dilution by metal-poor inflows. We find $\langle Z_{\star}\rangle_M/Z_\odot = 0.19\cdot\log(M_{\star}/M_\odot) - 1.27$ compared to \cite{Breda_2018}'s $\langle Z_{\star}\rangle_M/Z_\odot = 0.21\cdot\log(M_{\star}/M_\odot) - 1.52$ for disks. Since both works adopt consistent SSP libraries and metallicity binning, this modest offset is naturally interpreted as the effect of mild chemical enrichment over the $3-4$ Gyr look-back time between $z\sim0.3$ and $z=0$ and by the reduced extent of our sample and its scatter, compared to that in the local Universe.

The age gradients for the outer regions, presented in the same way as for the inner regions (Section \ref{sec:inner}), compare the mass and light-weighted stellar age gradients derived in this work (with \texttt{Starlight}) with those from \cite{Breda_2020} in the local Universe. In contrast to the inner regions' age gradients, the outer mass-weighted gradients ($R > R_{\rm eff}$; top panel) agree better, with $m = 0.11$, similar to \cite{Breda_2020} ($m = -0.26$ for $R_{\rm B}$ and $m = -0.27$ for $R_{\rm eff}$), with intercepts of $2$, $1.46$, and $-2.4$ repectively. The gradients remain close to $\nabla(t_{\star})_M \approx -1.2\,{\rm Gyr}/R_{\rm eff}$ across the full stellar mass range, consistent with the outer regions being dominated by younger stellar populations formed during extended disk growth. Finally, the outer luminosity-weighted gradients (bottom panel) exhibit similar overall trends to the local sample but are systematically less steep. While \cite{Breda_2020} find slopes of $m = -0.72$ for both $R_{\rm B}$ and $m = -0.95$ for $R_{\rm eff}$, our best-fit slope is $m = -0.6$ with intrceots of $6.3$,$8$ and $5.4$, respectively. The gradients range from $\nabla(t_{\star})_L \sim 0$ to $-1\,{\rm Gyr}/R_{\rm eff}$ at intermediate stellar masses, reaching values of $\sim -2\,{\rm Gyr}/R_{\rm eff}$ at the high-mass end. On average, the local gradients are steeper by approximately $1\,{\rm Gyr}/R_{\rm eff}$. This flattening of radial profiles at $z \sim 0.3$ could reflect more spatially extended recent SF, which contributes to younger, luminous stellar populations throughout the galaxies' outskirts. Additionally, the limited spatial coverage and physical resolution at this redshift may bias our measured gradients towards smaller absolute values.

\subsubsection{Outer regions: Star formation}
\label{sec::ssf_out}
\begin{figure}
    \centering
    \includegraphics[scale = 0.5]{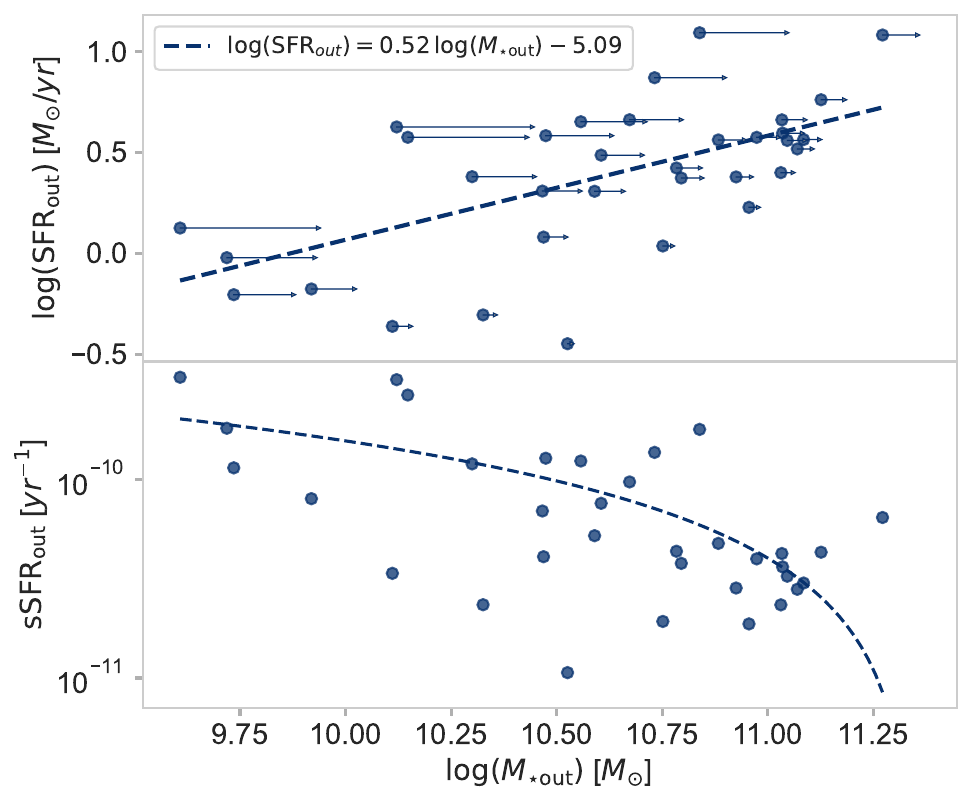}
    \caption{Top panel: Logarithmic SFR as a function of component stellar mass for outer regions. Arrows indicate the stellar mass that would be formed if the current SFR were constant between $z\sim0.3$ and $z=0$ with no mass loss, illustrating the potential impact of ongoing SF on component growth. Bottom panel: Specific star formation rate (sSFR) as a function of component stellar mass, fitted on a logarithmic scale. The properties presented are the result of this work with \texttt{Starlight}.}
    \label{fig:sfr_out}
\end{figure}
Similarly to Section \ref{sec::ssf_in}, the SF properties of the outer regions are shown in Fig.~\ref{fig:sfr_out}. The outer regions span a broad range of SFRs, with typical values of $\log(\mathrm{SFR}_{\rm out}) \sim 0.1-1\,[M_\odot\,\mathrm{yr}^{-1}]$, and display a clear positive correlation between SFR and component stellar mass. This trend is well described by a linear relation in logarithmic space, shown by the dashed line in the top panel of Fig.~\ref{fig:sfr_out}, indicating that the SFR of the outer regions increases steadily with increasing component stellar mass. The scatter around the best-fitting relation remains significant across the full mass range, reflecting a diversity of SF levels at a fixed mass, although not as pronounced as for inner regions.

Arrows in the Fig.~indicate the stellar mass that would be formed if the current SFR were sustained over the corresponding look-back time. The inferred accreted masses are substantial for most systems and increase towards lower stellar masses, suggesting that ongoing SF represents a non-negligible contribution to the stellar-mass budget of the outer regions within the explored mass range.

In the bottom panel of Fig.~\ref{fig:sfr_out}, the sSFR exhibits a systematic decline with increasing component mass, while remaining relatively high across the full sample. Quantitatively, $\log(\mathrm{sSFR}_{\rm out}/\mathrm{yr}^{-1})$ decreases from $\sim -9.3$ at $\log(M_{\rm out}/M_\odot)\sim10$ to $\sim -10.2$ at $\log(M_{\rm out}/M_\odot)\sim11.3$. Despite this decline, the outer regions maintain elevated sSFRs at all masses, indicating sustained SF activity relative to their assembled stellar mass.

\section{Discussion}
\label{discussion}
Our analysis paints a picture of galaxy evolution in which stellar assembly and quenching proceed from the inside out. Stellar population gradients across our $z\sim0.3$ galaxies are pronounced: both mass-weighted and luminosity-weighted ages decrease with radius, as do stellar metallicities. Central regions ($\rm R \leq 1\, \rm R_{\rm eff}$) uniformly host older ($t_{\star} \sim 5-10$ Gyr) and more metal-rich populations compared to their outskirts, which are younger and less metal-enriched (Sections \ref{sec:ages},\ref{sec:met}). We find the inner age gradients steepen within $\sim 1 \,R_{\rm eff}$, then flatten beyond $\sim 1.5 - 2\, R_{\rm eff}$, suggesting that once the central regions ceased forming stars, the outer disk became the dominant growth location (Sections \ref{sec:inner}, \ref{sec:outer}). 

Our findings on age profiles are broadly consistent with results from the PHANGS-MUSE survey \citep{Emsellem_2022} by \cite{Pessa_2023}, who investigated resolved stellar populations in nearby spiral galaxies. While their sample shows similarly negative gradients, our luminosity-weighted gradients appear generally steeper. This may indicate that galaxies at intermediate redshift were still actively building up their disks, with younger stellar populations in the outer regions, consistent with cosmological predictions of delayed SF in low-density outskirts \citep{Tacchella2015}.

Similar metallicity trends have been observed in local integral-field spectroscopy surveys such as CALIFA \citep{SanchezBlazquez2014, GonzalezDelgado2015}, MaNGA \citep{Parikh_2021}, and PHANGS-MUSE \citep{Pessa_2023}, which all report negative metallicity gradients that are mass- and structure-dependent. In the PHANGS-MUSE analysis, for instance, the stellar metallicity profiles show similar inside-out patterns and a dependence on morphology, with flatter gradients in barred or bulge-dominated systems. 
Such a result could be consistent with an inside-out growth scenario in which SF migrates outwards over time, as predicted by cosmological hydrodynamical simulations \citep{Avila_Reese_2018}, and chemical evolution models of disk formation \citep{Pilkington_2012}.
Recent results on gas-phase metallicity gradients provide an important complementary perspective. Using MAGPI galaxies at $z\sim0.35$, \cite{Mai_2026} find that gas-phase metallicity gradients are, on average, shallow, with a large diversity of profiles, including mostly flat and negative slopes. They show that this behaviour is closely linked to gas turbulence and mixing processes, driven by stellar feedback, gas accretion, and internal gas flows.
In this context, the mildly negative and, in some cases, nearly flat stellar metallicity gradients we observe suggest that both stellar and gas-phase metallicity distributions can appear similarly shallow, but for different physical reasons. While gas-phase metallicities are continuously reshaped by short-timescale processes such as mixing and inflows, stellar metallicities trace the cumulative record of past enrichment. 

We also observe a clear signature of inside-out quenching in the distribution of H$\alpha$ equivalent width, as a proxy for the sSFR (Section \ref{sec:ew}) \citep{Ellison_2018, Belfiore_2019}. EW$_{H\alpha}$) is highest in the outer regions (often $20-40$  \AA) and drops precipitously towards the centre, to values $\leq 5$  \AA\ in inner regions. High EW$_{H\alpha}$) indicates that young, high-mass stars dominate the local luminosity, whereas low EW can signify that SF has declined or ceased \citep{Belfiore2017}. This interpretation is further supported by recent spatially resolved studies of SF in intermediate-redshift galaxies. Using MAGPI data, \cite{Mun2024, Mun_2025} analysed SFR radial profiles and found that galaxies below the star-forming main sequence exhibit positive radial gradients, indicative of centrally suppressed SF, and thus inside-out quenching. In contrast, galaxies on or above the main sequence show flat or mildly positive profiles, consistent with more uniformly distributed SF across the disk. As a result, quenching processes act preferentially in the central regions, likely driven by internal mechanisms such as AGN feedback or reduced SF efficiency. The radial increase in EW$_{H\alpha}$ observed in our sample likewise reflects an increasing contribution of recent SF relative to the underlying stellar population at larger radii. 

Overall, stellar mass is the primary driver of the local age and metallicity of a stellar population, but the location within a galaxy and the current star-forming state modulate these properties in important ways. In our $z\sim0.3$ sample, massive inner parts ($\log (M_{\star \rm in}/M_{\odot})\geq10.5$) are a few Gyr older on average than the corresponding outer parts (Section \ref{sec:inner}). Outer regions yield a positive age-mass relation as well: more massive components (with $\log (M_{\star\rm out}/M_{\odot})\sim10.5-11$) reach ages of $>6$ Gyr, while lower-mass outskirts exhibit a broad range of younger ages ($\sim2-6$ Gyr) consistent with extended SF continuing to the present (Section \ref{sec:outer}). Strikingly, the slopes and shapes of these age-mass relations are very similar to those measured for the bulges and disks of local-Universe galaxies, the main difference being an overall shift towards younger ages at $z\sim0.3$. For a given mass, the median age at $z=0.3$ is roughly 3-4 Gyr lower than at $z=0$, which is the look-back time to $z=0.3$. 

Although the spectroscopic study of \cite{Damjanov_2025} focuses on quiescent galaxies, their results provide a useful reference for old stellar populations at intermediate redshift. They report mass-weighted stellar ages between $6.5$ and $10$ Gyr for massive ($\log(M_\star/M_\odot)\geq10.5$) quiescent systems at $0.3<z<0.6$. These ages are comparable to those we infer for the inner regions of the most massive galaxies in our star-forming sample, once differences in look-back time are accounted for, suggesting that their central stellar components already host old, quiescent populations while ongoing SF is mostly confined to larger radii. \cite{Jin_2024} also retrieved mass-weighted ages of bulges and disks for CALIFA spiral galaxies and found that bulges are systematically older than disk galaxies, with the bulge-disk age difference increasing with stellar mass.

As for metallicities, the typical inner-outer region contrast at high mass appears to be slightly more prominent than in less massive systems, with average $\Delta \log Z_{\rm in}\sim0.14$ and $\Delta \log Z_{\rm out}\sim0.11 $, respectively.
Consistent with these trends, spectroscopic studies of quiescent massive galaxies at comparable epochs report near-solar and sub-solar metallicities \citep{Damjanov_2025}. Although these studies probe a somewhat different galaxy population, the similarly high metallicities inferred for the inner regions of our most massive star-forming galaxies suggest that central stellar components can reach chemical maturity early, even while SF persists at larger radii. This behaviour is in line with the framework of cosmic downsizing \citep{Rigby_2015}, where the most massive galaxies enrich rapidly and complete their chemical evolution earlier than lower-mass systems.

To complement the evolutionary picture drawn from radial profiles and component mass properties, we find that outer regions currently form stars at significantly higher rates than inner regions of the same galaxy, even when the inner region contains a substantial fraction of the stellar mass (Sections \ref{sec::ssf_in},\ref{sec::ssf_out}). In fact, across our mass range, the specific SFR of outer regions is, on average, a factor of a few higher than that of inner ones. In lower-mass galaxies ($\log (M_{\star}/M_{\odot})\sim10$), the contrast is less extreme (sSFR $\sim3\times10^{-10}$ vs disk $\sim6\times10^{-10}$  yr$^{-1}$), but still the outskirts are forming stars more efficiently relative to their mass. This systematic central suppression of sSFR with increasing mass is a direct manifestation of downsizing in resolved form: high-mass galaxies not only have older bulges, but those bulges are allegedly contributing very little to stellar growth, whereas in lower mass galaxies, the bulges (which are typically smaller) can still participate in SF to some degree.
We note that \cite{Margalef_Bentabol_2017}, studying bulge and disk SFRs and sSFRs at $1<z<3$ in the CANDELS survey, reported a similar phenomenon: disk components maintained higher sSFR than bulges at all redshifts, and the sSFR gap widened towards lower $z$ as bulges quenched earlier. 

\subsection{Star formation histories}
\label{sec::sfh}
\begin{figure} 
    \centering
    \includegraphics[scale =0.45]{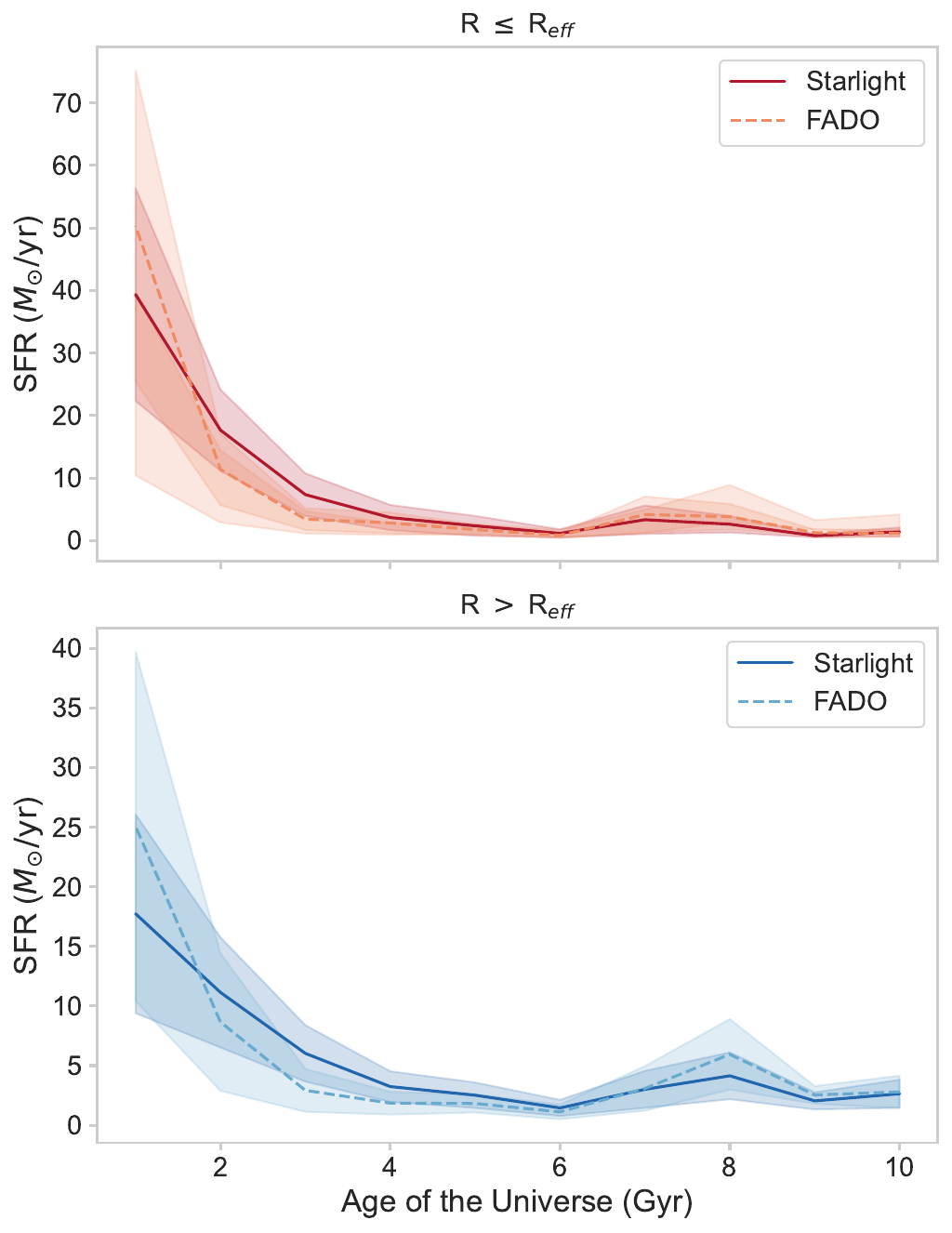}
    \caption{Star formation histories (SFHs) of inner (top) and outer (bottom) regions, averaged over our entire sample. Solid and dashed lines show the mean SFH obtained by averaging the contributions across the sample, while shaded regions refer to 1$\sigma$ values. Results are shown for both \texttt{FADO} (orange-light blue) and \texttt{Starlight} (red-dark blue) analysis.}
    \label{fig:sfh_components}
\end{figure}
To understand the temporal buildup of stellar mass in inner and outer regions, we analysed the average SFHs and the times at which each component assembled a given fraction of its total mass (denoted $\tau_{40}$, $\tau_{60}$, and $\tau_{80}$). The average SFHs shown in Fig.~\ref{fig:sfh_components} reveal a contrast between our inner and outer galactic regions. In each time bin, the SFH is computed from the mass assembly histories of individual galaxies as the total stellar mass ever formed within that time bin (depending on the chosen SSP library). Inner components are characterised by a rapid and intense early phase of SF, with the bulk of their stellar mass assembled within the first $1-2$ Gyr after the Big Bang. The outer regions show a more extended SFH. Although their SF activity also peaks at early epochs, they continue to form stars for several gigayears. The coexistence of an early peak with extended late-time SF also aligns with the trends reported by \cite{ScholzDiaz_2023}, who find that galaxies across a wide range of halo masses exhibit an early phase of stellar mass buildup followed by varying degrees of lower, continued SF.

An effect that is worth emphasising when interpreting the results of Fig.~\ref{fig:sfh_components} is that the SFHs derived with \texttt{FADO} tend to show a stronger central SF peak than the purely stellar \texttt{Starlight} fits, reflecting the different treatment of nebular emission and young stellar populations. Finally, we caution that part of the apparent SF excess that is noticeable both in the inner and outer regions might not be completely related to physical processes \citep{Conroy_2013}. It might arise from intrinsic limitations and artefacts of stellar population synthesis techniques. \cite{Cardoso_2022} (testing both \texttt{FADO} and \texttt{Starlight}) demonstrate that purely stellar spectral synthesis codes (that lack a self-consistent modelling of stellar and nebular emission) tend to overestimate the light or mass contributions of very young simple stellar populations $M_{\star} \leq 10^{7-9}M_{\odot}$, especially in galaxies with high sSFR or large nebular emission, when nebular continuum and line emission are neglected or insufficiently modelled. We therefore interpret any strong central recent SF activity excess with caution, considering that our fits, especially in inner bins, might include a contribution from modelling bias in addition to any real physical starburst episode.

While our interpretation is based on SFHs in different galactic regions, we note that secular evolution processes can substantially redistribute stellar populations over cosmic time. In particular, non-axisymmetric structures such as bars and spiral arms might induce radial migration, allowing stars to move from their birth radii while preserving relatively cold orbital kinematics \citep{Sellwood_2002, Roskar_2008}. Consequently, the present-day radial distribution of stellar populations does not necessarily trace the original spatial distribution of SF.

In addition, galaxy mergers, from repeated minor mergers with dwarf satellites to major merger events, can substantially redistribute stellar populations and alter disk structure. Although recent simulations suggest that stellar disks may survive or rapidly reform even after major merger events \citep{Sotillo_2022, Wu_2025}.
Recent theoretical work has further emphasised that transient spiral structure and collective secular processes can drive substantial stellar redistribution and structural evolution over time, complicating the interpretation of spatially resolved SFHs inferred from present-day stellar populations \citep{vdWel_2025, Bernaldez_2026, Meidt_2026, Minchev_2026}. Nevertheless, at this time our data do not allow us to robustly constrain the impact of these secular processes.

\subsection{Component mass fractions}
\begin{figure} 
    \centering
    \includegraphics[scale =0.475]{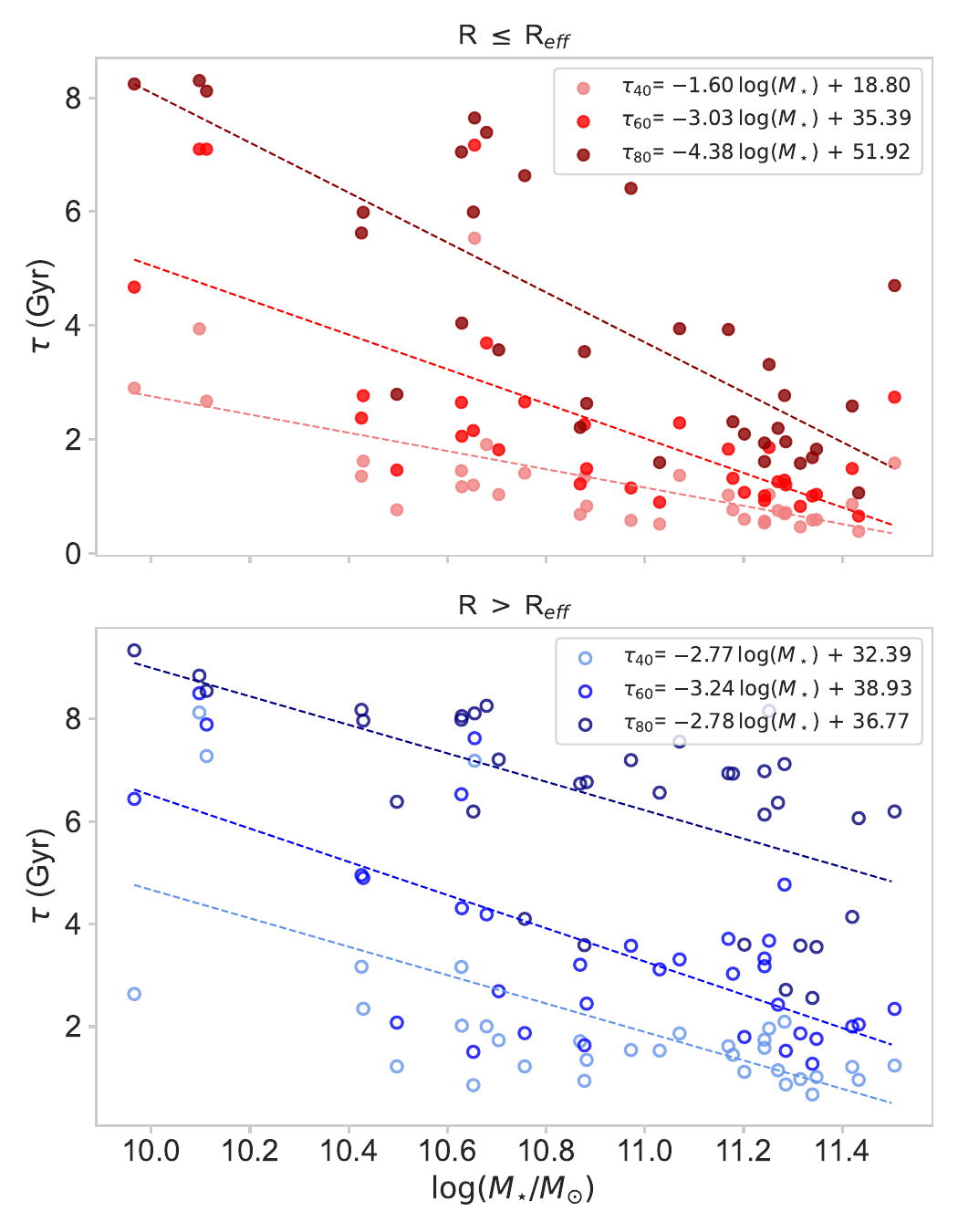}
    \caption{Characteristic assembly times when 40\%, 60\%, and 80\% of the stellar mass was assembled in inner and outer regions (in the top and bottom panels, respectively), as a function of stellar mass. The $\tau$ are colour-coded with blue and red colour maps: darker colours correspond to larger mass assembly fractions, respectively, $\tau_{40}$, $\tau_{60}$, and $\tau_{80}$. The dashed lines represent the best-fitting model to the data ($\tau = a\cdot \log(M_{\star}/M_{\odot})+b$), in the representative colours. The best-fitting parameter equations are reported in the Fig.~legend.}
    \label{fig:taus}
\end{figure}
The component mass fractions and assembly timescales quantify how the inner and outer regions contribute to the overall stellar mass growth of galaxies. Fig.~\ref{fig:taus} shows the look-back times at which 40\%, 60\%, and 80\% of the ever-formed stellar mass were assembled ($\tau_{40}$, $\tau_{60}$, and $\tau_{80}$, respectively). These time indices offer a robust description of the SF history, allowing for a direct comparison between components and across stellar-mass regimes. $\tau$ times are measured separately for the inner ($R \leq R_{\rm eff}$) and outer ($R > R_{\rm eff}$) regions and are derived from normalised cumulative mass-assembly curves reconstructed from the spectral synthesis analysis. The characteristic times $\tau_{40}$, $\tau_{60}$, and $\tau_{80}$ are taken as the times at which the cumulative mass fraction reaches 0.4, 0.6, and 0.8. An equivalent reconstruction was performed using both \texttt{Starlight} and \texttt{FADO}; while the qualitative trends are comparable between the two codes (see Appendix \ref{app:FS}), the results presented here correspond to \texttt{Starlight} outputs. 

A systematic offset between inner and outer regions is evident. At a fixed galaxy mass, the inner components consistently assemble earlier than the outer ones. The median $\tau_{40,\rm in}$ and $\tau_{60,\rm in}$ values occur $\sim1-2$ Gyr earlier than their outer counterparts, and the contrast increases with lower galaxy mass. The relations between assembly times and stellar mass are well described by linear fits, in the form $\tau = a \, \log(M_\star/M_\odot) + b$, whose coefficients are reported in Fig.~\ref{fig:taus}. More massive systems reach all $\tau$ thresholds at earlier cosmic epochs, reflecting the well-established downsizing behaviour observed in both global and spatially resolved studies \citep{Tacchella_2016a, Tacchella_2016b}. Specifically, inner regions in high-mass regimes typically achieve 80-90\% of their present-day stellar mass within the first $2-4\, \rm $ Gyr after the Big Bang, while the outskirts display more extended growth. The earliest assembly stages ($\tau_{40}$) correspond to the rapid, dissipative phase of SF, which accounts for the bulk of the buildup in the inner regions. Intermediate stages ($\tau_{60}$) mark the transition from centrally concentrated to spatially distributed SF, while the late-time index ($\tau_{80}$) is most sensitive to extended or rejuvenated growth of outer regions. The systematically delayed $\tau_{80,\rm out}$ values thus reinforce the picture of prolonged outskirts SF as the main driver of the recent increase in stellar mass.

In the inner parts, the differences between $\tau_{40}$, $\tau_{60}$, and $\tau_{80}$ are relatively small for higher masses (typically within $\sim1-1.5$ Gyr), indicating a rapid, nearly coeval buildup of stellar mass. This compressed timescale suggests that most of the central mass was formed during a short, intense formation episode. However, it increases significantly to 6 Gyr when we consider the lower-mass end of our sample. An additional trend is that the scatter in $\tau_{40}$ for the inner regions is considerably tighter than for $\tau_{48}$. This indicates that galaxies share relatively self-similar SFHs at early times but diverge more strongly at later epochs. In contrast, the outer components display much wider separations between the three $\tau$ thresholds, up to $\sim3-4$ Gyr. The larger $\tau_{60}-\tau_{80}$ intervals imply that SF continued at a significant level well after the bulk of the mass was assembled, pointing to a prolonged or episodic growth phase. Moreover, the three distinct slopes (especially $\tau_{40}$ and $\tau_{80}$) do have similar coefficients, at different formation stages, and do not seem to be as affected by galaxy mass. 

Spectral synthesis techniques that neglect or under-model nebular emission can overestimate the contribution of young stellar populations \citep{Cardoso_2019, Cardoso_2022}, leading to artificially delayed $\tau $ values. These uncertainties, while not altering the qualitative trends, should be kept in mind when interpreting the absolute timescales of late mass assembly. The lower $\tau$ and sSFR values of the inner regions confirm that their stellar mass was largely established during short, intense episodes at early cosmic times, followed by efficient quenching. In contrast, the outer regions retain higher sSFR values and later $\tau_{80}$ values, reflecting sustained or rejuvenated SF over several gigayears.

\section{Summary and conclusions}
\label{summary}
In this work, we have presented a spatially resolved stellar population analysis of star-forming galaxies at intermediate redshift ($0.28 < z < 0.35$) observed with the Multi Unit Spectroscopic Explorer (MUSE) as part of the MAGPI survey. By combining integral-field spectroscopy with full spectral synthesis using the \texttt{FADO} and \texttt{Starlight} codes via \texttt{GLANCE} pipeline, and adopting the \textsc{isan} technique to derive morphology-preserving radial profiles, we have constructed a detailed view of the stellar population structure and assembly histories at these redshifts. This approach, which couples morphological fidelity with physically self-consistent spectral modelling, represents a significant step forwards in connecting global trends at high redshift with the spatially resolved processes observed in the local Universe. Our main findings can be summarised as follows:
\begin{itemize}
    \item[-] We find clear overall negative radial gradients in stellar age and almost flat ones in metallicity (especially in the innermost regions), consistent with an inside-out formation scenario. The inner regions are systematically older and more metal-rich than their outer counterparts, with typical age differences of 2-3 Gyr and metallicity drops of $\sim 0.1$ to $0.2$ dex across one effective radius. Beyond $\sim 1.5 R_{\mathrm{eff}}$, both gradients flatten, suggesting that the outermost regions evolve more uniformly due to radial mixing and extended SF. \\
    \item[-] A strong mass dependence is observed: more massive galaxies exhibit shallower gradients and significantly older, metal-rich inner regions whose specific SF activity is already largely suppressed. In contrast, intermediate-mass systems retain younger and more actively star-forming outer regions, indicating that the transition towards centrally quenched structures was already well established by $z  \sim 0.3$. \\
    \item[-] The H$\alpha$ equivalent-width profiles further support this interpretation, showing centrally suppressed emission and rising values towards the outskirts. This pattern is consistent with a gradual fading of SF from the inside out. The most massive systems show low central $EW(\mathrm{H}\alpha)$ values ($<20$  \AA), typical of quenched galaxies, while the outskirts still display enhanced emission up to about 
    $40$\AA, pointing to ongoing or rejuvenated SF. \\
    \item[-] The reconstructed SFHs reveal distinct temporal behaviours between inner and outer regions. The inner parts formed the bulk ($\sim 80\%$) of their stellar mass rapidly within the first 2-3 Gyr of cosmic time, whereas the outskirts experienced more prolonged and gradual mass assembly, maintaining SF until later epochs. \\
    \item[-] The mass assembly times ($\tau$) highlight this dichotomy and show a strong dependence on galaxy mass. Inner regions show early, steep growth followed by more gradual evolution, whereas outer regions exhibit smoother and extended build-up histories. These patterns mirror the predictions of cosmological simulations and indicate that the mechanisms driving stellar mass assembly and quenching were already well established at intermediate redshift.
\end{itemize}
This study combines integral-field spectroscopy, full spectral synthesis including both stellar and nebular continuum emission, and morphology-preserving spatial sampling. The consistent results obtained from both \texttt{FADO} and \texttt{Starlight} (Appendix \ref{app:FS}) confirm the robustness of our approach and emphasise the advantages of spectroscopic analyses over purely photometric decompositions at these redshifts. Overall, this work demonstrates the diagnostic power of spatially resolved spectroscopy at intermediate redshift, capturing galaxies during a key transitional epoch between global SF and centrally quenched configurations.

\begin{acknowledgements}
We thank the ESO staff, and in particular the staff at Paranal Observatory, for carrying out the MAGPI observations. MAGPI targets were selected from GAMA. GAMA is a joint European-Australasian project based around a spectroscopic campaign using the Anglo-Australian Telescope. GAMA was funded by the STFC (UK), the ARC (Australia), the AAO, and the participating institutions. GAMA photometry is based on observations made with ESO Telescopes at the La Silla Paranal Observatory under programme ID 179.A-2004, ID 177.A-3016. The MAGPI team acknowledge support from the Australian Research Council Centre of Excellence for All Sky Astrophysics in 3 Dimensions (ASTRO 3D), through project number CE170100013. 
P.P. acknowledges support by Funda\c{c}\~{a}o para a Ci\^{e}ncia e a Tecnologia (FCT) grants UID/FIS/04434/2019, UIDB/04434/2020, UIDP/04434/2020 and Principal Investigator contract CIAAUP-092023-CTTI.
K.G. and T.N. acknowledge support from Australian Research Council Laureate Fellowship FL180100060. CF is the recipient of an Australian Research Council Future Fellowship (project number FT210100168) funded by the Australian Government. CF is a recipient of funding from the Australian Research Council (ARC) Discovery Project DP210101945.
GS acknowledge funding from the European Union's Horizon 2020 research and innovation programme under the Marie Sklodowska-Curie Grant agreement ID No.: 101147719.
KH acknowledges support by the Royal Society through a Dorothy Hodgkin Fellowship to KA Oman (DHF/R1/231105).
LMV acknowledges support by the German Academic Scholarship Foundation (Studienstiftung des deutschen Volkes) and the Marianne-Plehn-Program of the Elite Network of Bavaria.
SMS acknowledges funding from the Australian Research Council (DE220100003).
Parts of this research were conducted by the Australian Research Council Centre of Excellence for All Sky Astrophysics in 3 Dimensions (ASTRO 3D), through project number CE170100013.
SGL acknowledges the financial support from the MICIU with funding from the European Union NextGenerationEU and Generalitat Valenciana in the call Programa de Planes Complementarios de I+D+i (PRTR 2022) Project (VAL-JPAS), reference ASFAE/2022/025.
SLG is part of the research Project PID2023-149420NB-I00 funded by MICIU/AEI/10.13039/501100011033 and by ERDF/EU.
SLG is also supported by the project of excellence PROMETEO CIPROM/2023/21 of the Conselleria de Educación, Universidades y Empleo (Generalitat Valenciana)
\end{acknowledgements}

\bibliographystyle{aa}
\bibliography{ref}

\begin{appendix}

\section{\textsc{isan} radial dependencies}
\label{sec::radial}
\begin{figure*} 
    \centering
    \includegraphics[scale = 0.4]{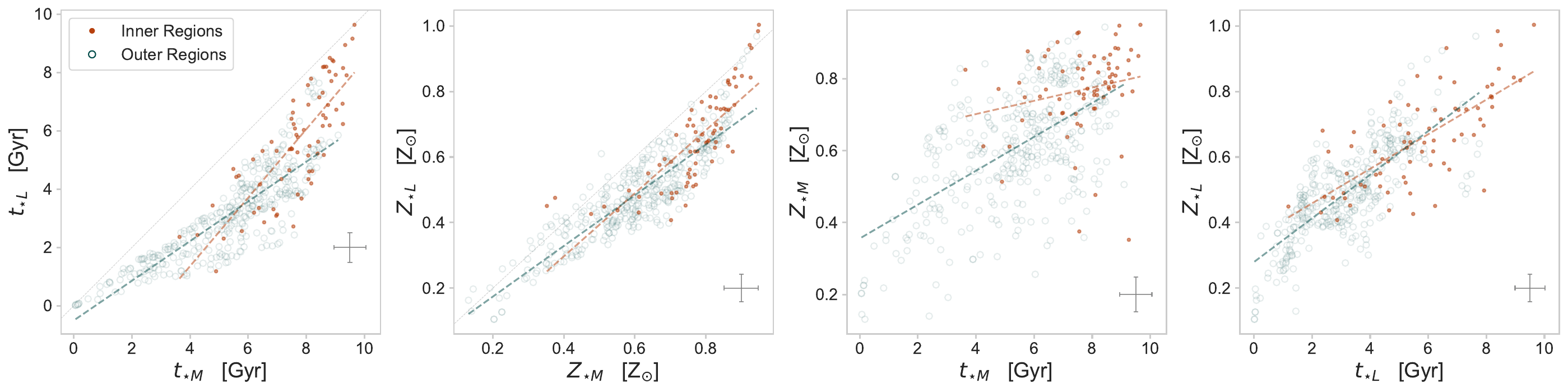}
    \caption{Comparison of mass-weighted vs luminosity-weighted stellar ages and metallicities, computed from inner (dots) and outer (circles) regions averages over isophotal annuli. For direct measurements of ages and metallicities, the dotted grey line is the bisector.}
    \label{fig:tz_bd_compare}
\end{figure*}
\begin{figure*} 
    \centering
    \includegraphics[scale = 0.4]{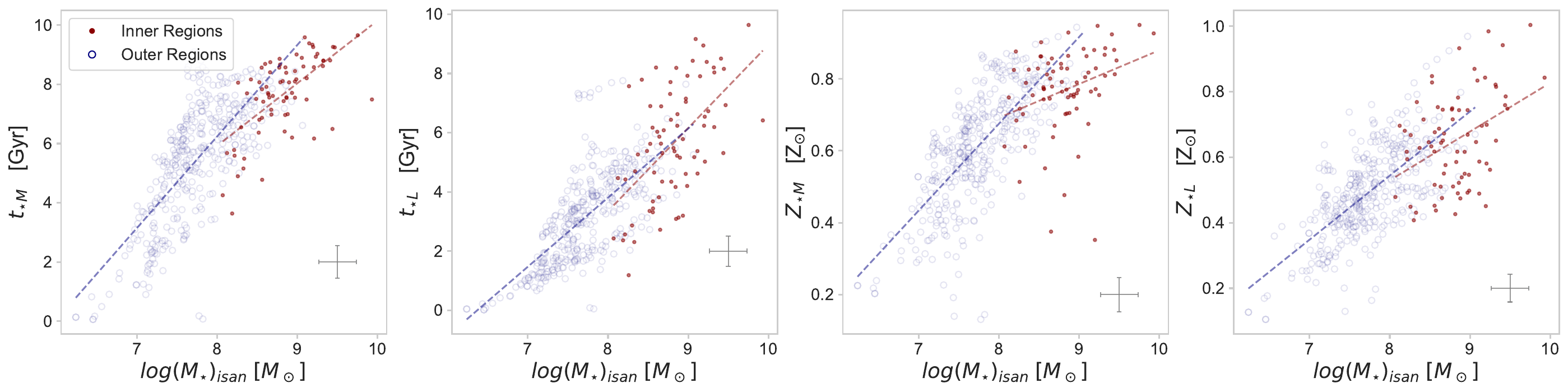}
    \caption{Mass-weighted and luminosity-weighted ages (top) and metallicities (bottom) plotted as a function of stellar mass ($\log(M_{\star})_{isan}$) in every isophotal annulus, for inner (dots) and outer (circles) components separately.}
    \label{fig:tz_mass_dep}
\end{figure*}
\begin{figure*} 
    \centering
    \includegraphics[scale = 0.4]{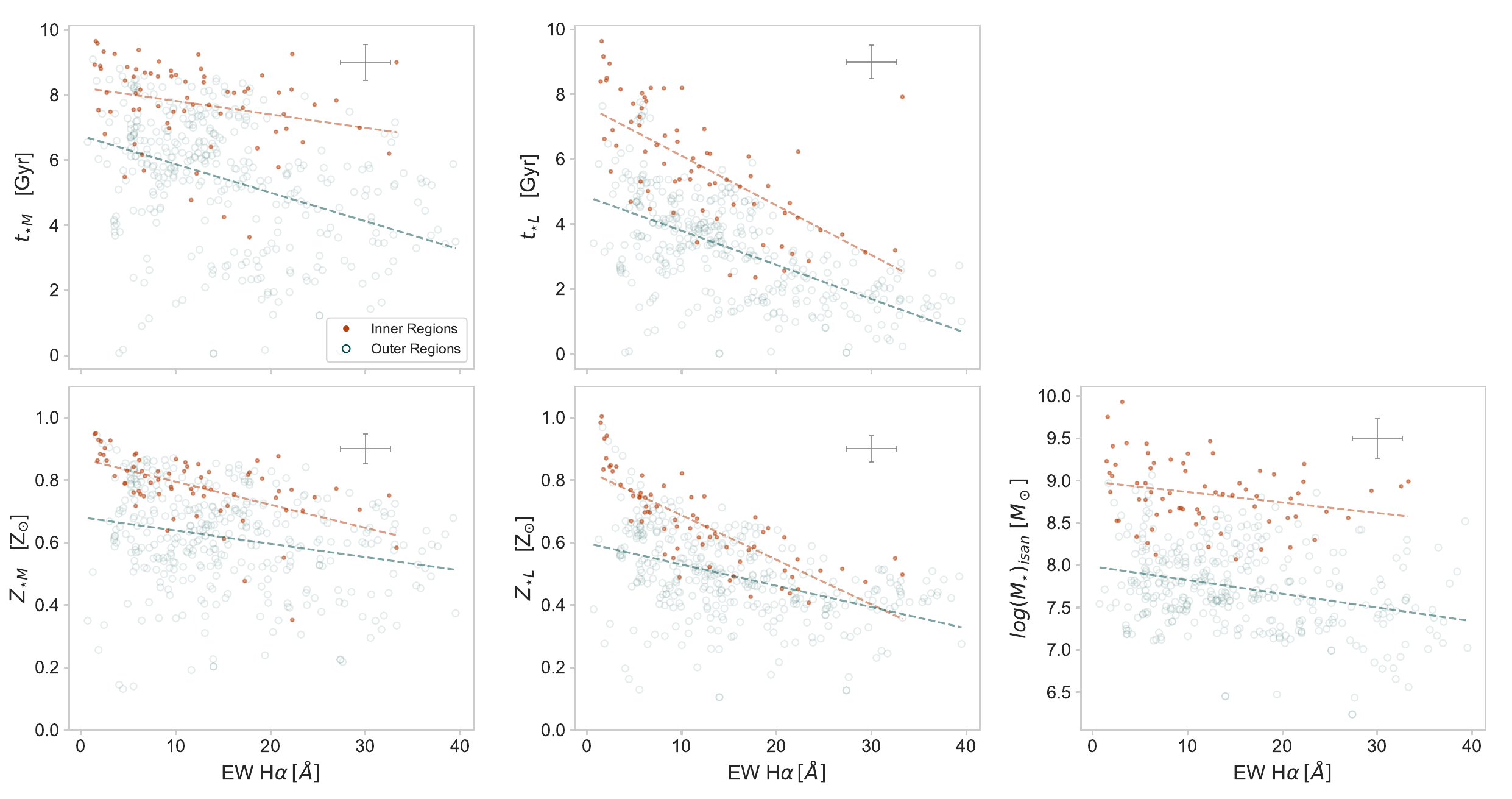}
    \caption{Correlation between $EW_{H\alpha}$ and stellar population properties (mass weighted and luminosity weighted ages and metallicities and mass), for inner (dots) and outer (circles) regions annuli separately.}
    \label{fig:ew_vs_stars}
\end{figure*}
Having distinguished the stellar populations' inner and outer isophotal zones, we now explore how their physical properties vary radially and depend on the adopted weighting scheme. In particular, we compare the mass and luminosity weighted ages and metallicities across isophotal annuli, and examine how these quantities scale with each other.

Fig.~ \ref{fig:tz_bd_compare} shows the relation between mass-weighted and luminosity-weighted ages and metallicities, averaged within inner and outer isophotal annuli. The top-left panel displays a tight correlation between $t_{\star \, \rm M}$ and $t_{\star \, \rm L}$, especially in inner regions, closer to the 1:1 line at $t\sim9-10$ Gyr. These uniformly old stellar populations likely formed rapidly at early epochs ($z\sim2$) and quenched on short timescales. By contrast, outer regions exhibit a broader distribution, with systematically younger luminosity-weighted ages ($t_{\star \, \rm L}\sim3-6$ Gyr) at $t_{\star \, \rm M}$ ($\sim 6-9$ Gyr). Such an age gap ($t_{\star \, \rm M} - t_{\star \, \rm L}$) demonstrates a transition from old, coeval stellar populations in the central regions to composite populations in the outskirts, where a young stellar component dominates the luminosity while the bulk of stellar mass remains old. The gap between the fitted orange and green lines and the grey bisector can indicate a mild radial dependence. The overall slightly larger $\Delta t$ values in the outer regions indicate that a small fraction of young stars dominates the luminosity, while the bulk of stellar mass remains old, consistent with ongoing SF rather than discrete rejuvenation events.

The comparison between metallicities ($Z_{\star \, \rm M}$ and $Z_{\star \, \rm L}$, top-right panel) shows a similar behavior. Inner parts occupy the high-metallicity regime ($Z_{\star}\sim0.6-1.0$), consistent with rapid early enrichment, while outskirts are typically sub-solar ($Z_{\star}\sim0.3-0.6$). In several star-forming outer regions, we find $Z_{\star \, \rm L} > Z_{\star \, \rm M}$, indicating that recent metal-rich SF episodes can bias luminosity-weighted quantities upwards \citep{looser2024}. The lower panels of Fig.~\ref{fig:tz_bd_compare} display the internal stellar age-metallicity correlations under each weighting scheme. Mass-weighted values ($t_{\star \, \rm M}-Z_{\star \, \rm M}$) exhibit a clear positive trend, less pronounced in inner regions, while in outer \textsc{isan} zones, the relation is more dispersed, pointing to a slower, more stochastic build-up of metals over extended timescales. Luminosity-weighted relations ($t_{\star \, \rm L}-Z_{\star \, \rm L}$) are instead more consistent between the different components, even while exhibiting more scatter, especially for the innermost regions.

Fig.~ \ref{fig:tz_mass_dep} further illustrates how these stellar population parameters depend on local stellar mass, $\log M_{\star}$, within each isophotal zone. Both age and metallicity increase systematically with $\log M_*$. Mass-weighted ages rise from $\sim \,3$ Gyr at $\log M_*\sim7$ to $>9$ Gyr at $\log M_{\star}\sim10$, forming a tight, monotonic sequence. Luminosity-weighted ages follow a similar trend but display larger scatter, especially in disk regions, where ongoing SF drives typical offsets of 2-3  Gyr between $t_{\star \, \rm M}$ and $t_{\star \, \rm L}$ at a fixed mass. Metallicity also correlates strongly with $\log M_{\star}$: $Z_{\star \, \rm M}$ increases steadily from $\sim \,0.2$ to $\sim \,0.9$  $z_\odot$, while $Z_{\star \, \rm L}$ values show enhanced dispersion at low mass, occasionally exceeding $Z_{\star \, \rm M}$ due to localised, metal-rich star-forming events. Inner regions dominate the high-mass, high-age, high-metallicity end of all relations, whereas outskirts span lower masses and display greater internal variation. 

To connect these stellar population trends with current SF activity, Fig.~\ref{fig:ew_vs_stars} shows how the derived parameters scale with the H$\alpha$ equivalent width, which traces the sSFR. As expected, EW$_{H\alpha}$ decreases systematically with increasing stellar age, with both $t_{\star \, \rm M}$ and $t_{\star \, \rm L}$ exhibiting clear anti-correlations. The dependence is steeper for $t_{\star \, \rm L}$, reflecting the dominance of young OB stars in the ionising continuum. Central regions with EW$_{H\alpha}<3$  \AA\ correspond to old, quiescent stellar populations ($t_{\star \, \rm L}>8$ Gyr), while outer zones reach EW$_{H\alpha}$$\sim20-30$  \AA, associated with much younger luminosity-weighted ages ($t_{\star \, \rm L}\sim3-5$  Gyr).

Metallicity shows an opposite trend: regions with low EW$_{H\alpha}$ tend to be more metal-rich, reflecting the cumulative enrichment of old stellar populations, while high-EW zones in the outskirts are relatively metal-poor. The effect is stronger for luminosity-weighted metallicities, as ongoing SF can temporarily elevate $Z_{\star \, \rm L}$ through the presence of young, metal-rich stars \citep{GonzalezDelgado2015, looser2024}. The correlation between stellar mass and EW$_{H\alpha}$ further reinforces this picture: massive central regions exhibit weak emission and old, enriched populations, whereas low-mass outer regions display strong emission and younger, metal-poor stars, consistent with an inside-out formation scenario \citep{Belfiore2017, Perez_2021}. 

Overall, the convergence of trends across both figures highlights a coherent evolutionary picture: central, dense regions assembled early and enriched rapidly, while outer regions continued forming stars over extended periods. The tighter correlations and lower scatter observed in mass-weighted quantities confirm that they are more faithful tracers of the integrated SF and chemical enrichment histories, whereas luminosity-weighted parameters are more sensitive to short-term SF episodes. This dichotomy reinforces the interpretation of inside-out galaxy growth and underscores the necessity of mass-weighted diagnostics for recovering the true evolutionary pathways of composite stellar systems.

\section{\texttt{FADO} versus \texttt{Starlight}}
\label{app:FS}
To assess the consistency between stellar population synthesis methods, we compare the key physical quantities derived using \texttt{FADO} and \texttt{Starlight}. The comparisons are done separately for inner and outer regions, and include both average total properties and values averaged in isophotal annuli.

\begin{figure*}
    \centering
    \includegraphics[scale = 0.4]{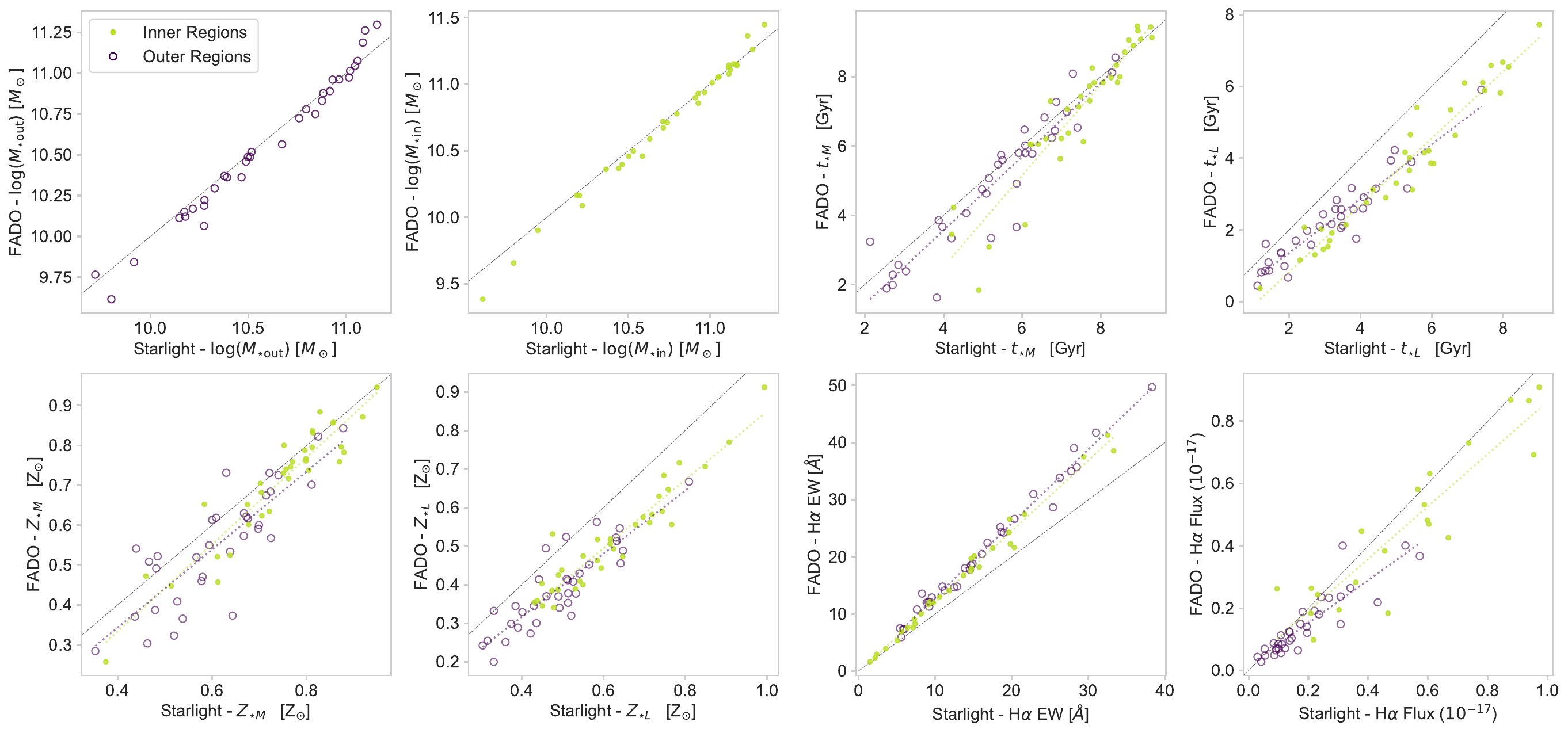}
    \caption{Comparison between \texttt{FADO} and \texttt{Starlight} for global inner regions (dots) and outer regions (circles) properties: stellar mass, mass-weighted and luminosity-weighted ages and metallicities, H$\alpha$ EW and flux. The dashed grey line indicates the 1:1 bisector.}
    \label{fig:BD_method_compare}
\end{figure*}
In Fig.~ \ref{fig:BD_method_compare}, we compare the average inner regions and outer regions values for both methods. There is overall good agreement, especially in stellar mass and mass-weighted properties. Luminosity-weighted values show increasing differences between the two methods, particularly in the inner regions, likely due to differences in how the emission lines are modelled.
H$\alpha$ flux and equivalent width show some scatter. This reflects their sensitivity to the treatment of emission lines: \texttt{FADO} includes nebular continuum and line modelling by design, while \texttt{Starlight} generally relies on absorption-dominated templates with emission treated separately.

\begin{figure*}
    \centering
    \includegraphics[scale = 0.4]{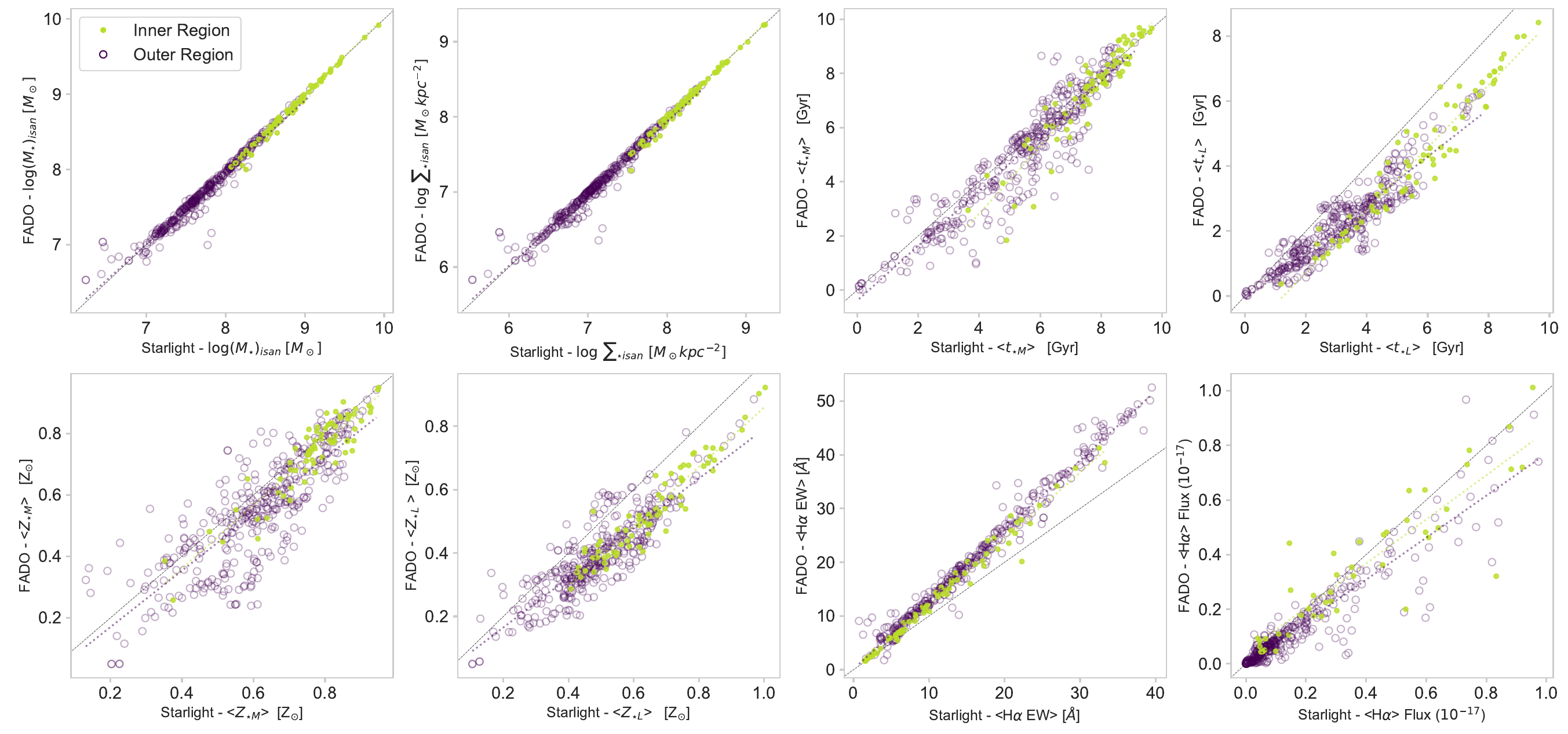}
    \caption{\texttt{FADO} vs \texttt{Starlight} comparison for quantities measured along isophotal annuli. Each point represents a radial value in outer (circles) and inner (dots) regions. The dashed grey line indicates the 1:1 bisector.}
    \label{fig:isoph_method_compare}
\end{figure*}
To assess the consistency between different stellar population synthesis approaches, we compare the main physical quantities derived with \texttt{FADO} and \texttt{Starlight}. Fig.~\ref{fig:BD_method_compare} shows the correspondence between global quantities for inner and outer components, while Fig.~\ref{fig:isoph_method_compare} presents the comparison for radially averaged values within isophotal annuli. In both cases, each panel contrasts the outputs of the two codes for stellar mass and density, mass and luminosity weighted age and metallicity, as well as H$\alpha$ flux and equivalent width. Inner and outer regions are plotted as filled and open symbols, respectively.

The two methods yield closely matching stellar masses and mass-weighted quantities, indicating that the inferred long-term SFHs are consistent. However, it's nevertheless important to note that luminosity-weighted quantities are generally more sensitive to small fractions of young stellar populations, since young stars dominate the emitted light even when contributing little to the total stellar mass. Mass-weighted quantities instead are dominated by the older stellar populations that contain most of the stellar mass, making them less sensitive to recent SF episodes. Consequently, mass-weighted quantities are typically more stable than luminosity-weighted ones.
The most noticeable scatter is in the stellar metallicities, both for mass-weighted and luminosity-weighted estimates. These differences likely arise from the treatment of nebular emission and recent SF episodes: \texttt{FADO} self-consistently models both the stellar and nebular continua, while \texttt{Starlight} typically fits purely stellar templates with emission lines masked or treated separately. This also explains why H$\alpha$ fluxes and equivalent widths show systematic deviations, particularly in regions with prominent emission.
Overall, the agreement remains tight for mass and age profiles, while metallicities and emission-line quantities display increasing scatter towards low-surface-brightness outer regions, where the S/N decreases, and nebular contributions become more uncertain. These tests provide an estimate of the intrinsic systematic differences expected when inter-comparing results based on distinct population synthesis codes, demonstrating that global and radial trends are overall robust despite methodological differences.

\end{appendix}
\end{document}